%% file: main.tex
\ifdefined\pdfobjcompresslevel
\fi
\documentclass[sigplan,nonacm]{acmart}

\input{pluverse-latex-style-guide/macro/pluverse-preamble.tex}
\hypersetup{hidelinks}

\newcommand{\sourceFunc}{\ensuremath{\mathit{src}}\xspace}
\newcommand{\targetFunc}{\ensuremath{\mathit{tgt}}\xspace}
\newcommand{\cx}{Cx\xspace}%
\newcommand{\compactUnderbrace}[2]{%
  \mathord{\vtop{\offinterlineskip\mathsurround=0pt
    \ialign{##\crcr
      $\hfil\displaystyle#1\hfil$\crcr
      \noalign{\kern1pt}
      \upbracefill\crcr
      \noalign{\kern1pt}
      $\hfil\scriptstyle#2\hfil$\crcr
    }%
  }}%
}

\lstdefinelanguage{llvm}{
  morekeywords={define, declare, ret, call, void, ptr, global},
  morekeywords=[2]{add, sub, mul, shl, lshr, ashr, and, or, xor, icmp, zext, sext, load, store, getelementptr, select, br, phi, trunc, isPowerOf2,ctpop},
  morekeywords=[3]{i1, i2, i4, i8, i16, i32, i64, half, float, double},
  morecomment=[l]{;},
  sensitive=true
}

\definecolor{L3blue}{HTML}{2A78D6}
\definecolor{L3orange}{HTML}{EB6834}
\input{constants}
\input{pluverse-latex-style-guide/macro/pluverse-preamble-end.tex}

\pgfplotsset{compat=1.18}
\begin{document}

\setlength{\abovedisplayskip}{4pt}
\setlength{\belowdisplayskip}{4pt}
\setlength{\abovedisplayshortskip}{4pt}
\setlength{\belowdisplayshortskip}{4pt}

\title{\llvm Translation Validation Automated with Large Language Models and \lean}

\author{Chunhao Liao}
\orcid{0009-0004-4379-9454}
\affiliation{
   \institution{University of Waterloo}
     \country{Canada}
}
\email{chunhao.liao@uwaterloo.ca}

\author{Hongxu Xu}
\orcid{0009-0003-0926-220X}
\affiliation{
   \institution{University of Waterloo}
     \country{Canada}
}
\email{hongxu.xu@uwaterloo.ca}

\author{Xintong Zhou}
\orcid{0009-0002-6444-5431}
\affiliation{%
	\institution{University of Waterloo}
	\country{Canada}}
\email{x27zhou@uwaterloo.ca}

\author{Yizhou Zhang}
\orcid{0000-0002-8206-4694}
\affiliation{%
	\institution{University of Waterloo}
	\country{Canada}}
\email{yizhou@uwaterloo.ca}

\author{Chengnian Sun}
\orcid{0000-0002-0862-2491}
\affiliation{
   \institution{University of Waterloo}
   \country{Canada}
}
\email{cnsun@uwaterloo.ca}

\begin{abstract}
\input{abstract}
\end{abstract}

\keywords{Compilers, Translation Validation, Large Language Models, Formal Verification}

\maketitle %

\section{Introduction}
\label{sec:intro}
\input{intro}

\section{Background}
\label{sec:background}
\input{background}

\section{Methodology}
\label{sec:methodology}

\input{methodology}

\section{Evaluation}
\label{sec:evaluation}
\input{evaluation}

\section{Related Work}
\label{sec:related-work}
\input{related-work}

\section{Conclusion}
\label{sec:conclusion}
\input{conclusion}

\bibliographystyle{ACM-Reference-Format}
\bibliography{acmart}

\newpage

\end{document}

%% file: pluverse-latex-style-guide/macro/pluverse-preamble.tex
\usepackage[ruled,lined,linesnumbered,vlined]{algorithm2e}

\SetCommentSty{mycommfont}
\DontPrintSemicolon % Do not show the semicolons.

\SetKwProg{Fn}{Function}{:}{}

\usepackage{amsmath}
\usepackage{amsfonts}
\usepackage{amsthm}
\usepackage{balance}
\usepackage{enumitem}
\usepackage{graphicx}
\usepackage{listings}
\usepackage{multicol}
\usepackage{multirow}
\usepackage{subcaption}
\usepackage{url}
\usepackage{xspace}
\usepackage{xcolor}
\usepackage{tcolorbox}
\usepackage{colortbl}
\usepackage{siunitx}
\usepackage{tikz}
\usepackage{pgfplots}
\newcommand*\circled[1]{\tikz[baseline=(char.base)]{
    \node[shape=circle,draw,inner sep=0.5pt] (char) {\small#1};}}
\usetikzlibrary{shapes}
\usetikzlibrary{shapes.geometric}
\usetikzlibrary{arrows.meta, positioning}

\newcommand{\eg}{\mbox{\textit{e.g.}}\xspace}

\newcommand{\ie}{\mbox{\textit{i.e.}}\xspace}
\definecolor{BlindColorTolOne}{HTML}{332288}
\definecolor{BlindColorTolTwo}{HTML}{117733} % green
\definecolor{BlindColorTolThree}{HTML}{44AA99}
\definecolor{BlindColorTolFour}{HTML}{88CCEE}
\definecolor{BlindColorTolFive}{HTML}{DDCC77}
\definecolor{BlindColorTolSix}{HTML}{CC6677} % light red
\definecolor{BlindColorTolSeven}{HTML}{AA4499}
\definecolor{BlindColorTolEight}{HTML}{882255}

\definecolor{BlindColorWongOne}{HTML}{000000} % black
\definecolor{BlindColorWongTwo}{HTML}{E69F00}
\definecolor{BlindColorWongThree}{HTML}{56B4E9}
\definecolor{BlindColorWongFour}{HTML}{009E73}
\definecolor{BlindColorWongFive}{HTML}{F0E442}
\definecolor{BlindColorWongSix}{HTML}{0072B2} % blue
\definecolor{BlindColorWongSeven}{HTML}{D55E00}
\definecolor{BlindColorWongEight}{HTML}{CC79A7}
\definecolor{mygreen}{HTML}{02818a}

\mathchardef\mhyphen="2D

\newcounter{FindingCounter}

\newcommand{\myparagraph}[1]{
  \vspace*{0.04cm}
  \noindent \textit{\textbf{#1.}}\quad
}

\newcommand{\itemParagraph}[1]{%
  \vspace*{0.04cm}\noindent\textit{#1.}\quad}

\newcommand{\projectname}[1]{\mbox{\textsf{#1}}\xspace} % \xspace is not needed here.
\SetKwData{AlgTrue}{true}
\SetKwData{AlgFalse}{false}
\SetKw{AlgBreak}{break}
\SetKw{AlgContinue}{continue}

\newcounter{myUniqueIdCounter}

\makeatletter
\newcommand{\myGetOrAssignID}[1]{%
  \ifcsname myMap@#1\endcsname%
    \csname myMap@#1\endcsname%
  \else%
    \stepcounter{myUniqueIdCounter}%
    \expandafter\xdef\csname myMap@#1\endcsname{\themyUniqueIdCounter}%
    \themyUniqueIdCounter%
  \fi%
}
\makeatother

\newcommand{\anonymizedId}[1]{\ifx\useAnonymizedId\undefined%
  #1%
\else%
  \myGetOrAssignID{#1}%
\fi}

%% file: pluverse-latex-style-guide/macro/pluverse-preamble-end.tex
\usepackage{cleveref} % This package must be loaded in the end.

\Crefname{algocf}{Algorithm}{Algorithms}
\crefname{algocf}{Algorithm}{Algorithms}

\Crefname{algorithm}{Algorithm}{Algorithms}
\crefname{algorithm}{Algorithm}{Algorithms}

\crefname{appendix}{Appendix}{Appendices}
\Crefname{appendix}{Appendix}{Appendices}

\Crefname{figure}{Figure}{Figures}
\crefname{figure}{Figure}{Figures}

\crefname{listing}{Listing}{Listings}
\Crefname{listing}{Listing}{Listings}

\Crefname{table}{Table}{Tables}
\crefname{table}{Table}{Tables}

\crefname{thm}{Theorem}{Theorems}
\Crefname{thm}{Theorem}{Theorems}

\crefname{definition}{Definition}{Definitions}
\Crefname{definition}{Definition}{Definitions}

\crefname{lemma}{Lemma}{Lemmas}
\Crefname{lemma}{Lemma}{Lemmas}

\crefname{corollary}{Corollary}{Corollaries}
\Crefname{corollary}{Corollary}{Corollaries}

\crefname{proof}{Proof}{Proofs}
\Crefname{proof}{Proof}{Proofs}

\crefname{equation}{Equation}{Equations}
\Crefname{equation}{Equation}{Equations}

\crefformat{chapter}{\S~#2#1#3}
\crefmultiformat{chapter}{\S\S~#2#1#3}{ and~#2#1#3}{, #2#1#3}{, and~#2#1#3}

\crefformat{section}{\S~#2#1#3}
\crefmultiformat{section}{\S\S~#2#1#3}{ and~#2#1#3}{, #2#1#3}{, and~#2#1#3}

%% file: abstract.tex
\llvm is the cornerstone of modern compilers, but its subtle intermediate
representation (\ir) semantics make transformations error-prone
and necessitate formal verification.
\alivetwo, a state-of-the-art translation validator based on
satisfiability modulo theories, has achieved substantial success
in automating the validation of \llvm transformations.
However, it still faces scalability limitations, does not support symbolic
bitwidths, and offers only bounded guarantees for loops.
In contrast, interactive theorem provers such as \lean can
address these cases but require substantial proof engineering.

In this paper, we present \proj, a framework combining large language models (\llms)
and \lean for automated translation validation of \llvm transformations.
\proj generates structured proof scaffolds based on source and target functions,
automatically discharges obligations amenable to deterministic reasoning,
and delegates transformation-specific obligations to \llms.
It produces refinement proofs or counterexample-based refutations,
with every successful verdict checked by the \lean kernel.
On \ValBenchCount \llvm transformations, \proj verifies or refutes
\ValSolvedCount, leaving one invalid case unresolved.
Successful cases include \ValParamBWSolvedCount loop-free transformations
with symbolic bitwidths, \ValLoopSolvedCount cases from a restricted class
of loop-containing transformations, and \ValFixedBWTimeoutCount complex
valid fixed-bitwidth cases on which \alivetwo times out.
Compared with an unscaffolded baseline, scaffolding enables
\ValScaffoldNewSolveCount additional proofs.
On cases solved by both configurations, it reduces mean proof time by
\ValScaffoldWallTimeReduction and mean monetary cost by
\ValScaffoldCostReduction.

%% file: intro.tex
The \llvm compiler infrastructure~\cite{LLVM} is a cornerstone of modern
software development in industry and academia.
At its core is the \llvm Intermediate Representation (\ir), a strongly typed
language that bridges high-level source languages and diverse target
architectures.
To enable aggressive optimization, the \ir provides highly expressive
semantics, including Undefined Behavior (\ub)~\cite{llvm_ub_manual}.
This expressiveness comes at a cost: combined with the sophisticated
reasoning required by optimization passes, it makes correct implementation
notoriously difficult---developers must account for every edge case by hand,
and a single overlooked one can silently produce a miscompilation.
Indeed, optimization passes account for a significant portion of reported
bugs in \llvm~\cite{yang2011finding,emi,issta2016compilerbug}, making
reliable optimizer development a persistent challenge.

\myparagraph{Translation Validation}
To provide formal
guarantees that compiler transformations preserve program semantics,
modern compiler development increasingly relies on translation validation~\cite{tv,opttv}.
This technique compares a source function
(pre-transformation program)
with its corresponding target function (post-transformation program)
using a formal semantic model, ultimately either proving that
the target refines the source (see \cref{def:refinement})
or generating a counterexample if the transformation is invalid.
A prominent instance of this methodology is \alivetwo~\cite{alivetwo}, a state-of-the-art, \smt-based~\cite{barrett-smtbookch21} \emph{bounded} translation validator specifically
designed for \llvm \ir. \alivetwo has been widely
adopted by the \llvm community to rigorously vet
compiler transformations before they are accepted~\cite{llvm_instcombine_contributor_guide}.

Despite its practical effectiveness, \alivetwo has three inherent limitations:
\begin{enumerate}[leftmargin=*]

\item \textbf{Solver Scalability.}
As an \smt-based tool, \alivetwo can time out on large bitwidths,
long instruction sequences, and complex
formulas~\cite{smtbitblasting}.

\item \textbf{Fixed-Bitwidth Reasoning.}
\alivetwo only supports validation of fixed-bitwidth transformation instances,
limiting its ability to verify optimizations that
hold across symbolic bitwidths~\cite{hydra,lpg}.

\item \textbf{Bounded Loop Unrolling.}
\alivetwo validates loops by unrolling them up to a fixed bound,
leaving subsequent iterations unchecked.
\end{enumerate}
These limitations motivate a complementary approach to formally
validating \llvm transformations.

Interactive theorem provers (\itps) offer a promising alternative for translation validation.
Compared to \smt-based bounded translation validation, an \itp
can efficiently express and validate transformations involving
symbolic bitwidths, large bitwidths, long instruction sequences,
and loops.
While these benefits are compelling in principle, the practical adoption of such provers has historically
been hindered by the prohibitive cost of proof engineering,
which often requires an order of magnitude more proof code than
implementation code~\cite{SquirrelFS,vMVCC,GoJournal}.
However, recent advances in Large Language Models (\llms) make \itp-based workflows increasingly viable. \llms can iteratively synthesize and repair proof scripts, while the \itp
kernel rigorously guarantees the correctness of the final proof~\cite{proofautomation,Rango,COPRA}.

\myparagraph{\proj}
To this end, this paper presents \proj\footnote{\proj vets LLVM transformations
based on the integration of three components: Lean, LLMs, and LLVM.},
a framework that pioneers the application of \llms and
\lean~\cite{lean,leanfour} to the automated
translation validation of \llvm transformations.
By leveraging domain knowledge of compiler optimizations, \proj
automatically extracts a \lean refinement theorem and structured
proof scaffolds directly from the source and target functions.
These scaffolds not only minimize expensive \llm queries, but also
systematically guide the \llm through both verification and
refutation.
For verification, the scaffold leaves focused typed holes for
transformation-specific proof obligations, which an \llm can solve,
using \lean's diagnostic feedback to iteratively repair proofs when
necessary. For refutation, a structured prompt guides the \llm
to propose a concrete counterexample. \proj uses a deterministic
scaffold to check the candidate counterexample and, upon
successful validation, construct a machine-checked proof of non-refinement.
In both cases, \lean remains the trusted checker of the final verdict.

We evaluate \proj on a dataset of
\ValBenchCount \llvm transformations. The dataset includes a loop-free subset
(\DsTotal valid, \DsMut invalid) and a loop-containing subset
(\ValLoopValidCount valid, \ValLoopInvalidCount invalid). Both subsets
cover fixed-bitwidth and symbolic-bitwidth cases.
Overall, \proj successfully validates all \ValVerifiedCount
valid transformations and refutes \ValRefutedCount invalid ones,
leaving only one invalid transformation unresolved.
Moreover, compared to an unscaffolded baseline, scaffolding
reduces mean proof time by \ValScaffoldWallTimeReduction and
mean monetary cost by \ValScaffoldCostReduction,
while enabling \ValScaffoldNewSolveCount additional proofs in total.
In contrast, \alivetwo validates
\AliveFixedBWVerifiedCount out of \AliveFixedBWCount loop-free, fixed-bitwidth
\llvm transformations, but cannot validate symbolic-bitwidth cases
or provide unbounded guarantees for loop-containing transformations.
These results demonstrate \proj's effectiveness in automating
the translation validation of \llvm transformations.

\noindent\textbf{\emph{Contributions.}} \quad
Our main contributions are listed below.
\begin{itemize}[leftmargin=*, topsep=0pt, partopsep=0pt]
    \item
    We introduce \proj, the first framework applying \llms and \lean
    to the automated translation validation of \llvm transformations:
    the \llm synthesizes refinement proofs or counterexamples,
    while the \lean kernel rigorously certifies every verdict.

    \item
    We develop an automated, non-\llm generator that extracts a \lean
    refinement theorem and structured proof scaffolds from the source
    and target functions. These scaffolds encode compiler domain
    knowledge to guide the \llm,
    reducing validation to focused, transformation-specific holes and
    converting \llm-proposed counterexamples into machine-checked proofs.

    \item
    We evaluate \proj on \ValBenchCount \llvm transformations, successfully
    validating all \ValVerifiedCount valid cases and refuting \ValRefutedCount
    invalid ones, leaving only one invalid transformation unresolved.
    \proj validates symbolic-bitwidth and loop-containing
    transformations beyond \alivetwo's scope. Scaffolding reduces mean
    proof time by \ValScaffoldWallTimeReduction and mean monetary cost by
    \ValScaffoldCostReduction compared to an unscaffolded baseline, enabling
    \ValScaffoldNewSolveCount additional proofs in total.
\end{itemize}

%% file: background.tex
This section provides necessary background on \llvm transformations
and \llvm translation validation.

\subsection{\llvm Transformations}
\label{subsec:llvm-transformations}

\llvm transformations rewrite \llvm \ir while preserving the program's
observable behavior. They range from local instruction-level
rewrites to optimizations involving multiple basic blocks or
loops. Following prior work~\cite{hydra},
we represent a transformation by the following judgment:
\[
\precondition \vDash \lhs \Rightarrow \rhs
\]
where $\lhs$ (left-hand side) denotes the pre-transformation fragment, $\rhs$
(right-hand side) denotes its replacement, and $\precondition$
is a logical predicate over the operands in $\lhs$
that defines when the transformation is valid.
Specifically, this judgment states that replacing
$\lhs$ with $\rhs$ is semantics-preserving
when $\precondition$ holds.
When no explicit precondition is specified, $\precondition$
defaults to $\AlgTrue$, indicating that the transformation
is always valid and imposing no constraints on the operands.

In practice, \llvm \ir translation validation instantiates this
judgment as a pair of \llvm functions~\cite{llvm_instcombine_contributor_guide}:
a \sourceFunc function, which encodes $\precondition$ and $\lhs$, and a
\targetFunc function, which encodes $\rhs$.
The $\precondition$ is expressed within the
\sourceFunc function through the \texttt{llvm.assume}
intrinsic,
which restricts the inputs to those satisfying
$\precondition$~\cite{llvm_langref_assume}.

\myparagraph{Undefined Behavior Semantics}
\llvm has two forms of Undefined Behavior (\ub):
\emph{immediate} \ub and \emph{deferred} \ub~\cite{llvm_ub_manual}.
Immediate \ub, such as division by zero, permits arbitrary program
behavior.
Deferred \ub, such as integer overflow, instead produces an
ill-defined value that propagates through downstream computations.
These ill-defined values take two forms in \llvm:
\emph{poison}~\cite{llvm_langref_poison} and
\emph{undef}~\cite{llvm_langref_undef}.
Our work supports \emph{poison},
which taints dependent computations and triggers immediate
\ub when used in a way that requires a well-defined value, such as a
branch condition.
In contrast, an \emph{undef} value nondeterministically takes
any value of its type at each use, which has long complicated
reasoning about \llvm semantics.
\llvm has deprecated
undef~\cite{remove_undef_values},
and thus our work
does not support it.

\myparagraph{Refinement Relation}
Since optimizers routinely exploit \ub, correctness cannot be formulated as equivalence.
Following \alivetwo~\cite{alivetwo,llvm_ub_refinement_lattice}, we formalize transformation correctness in terms of refinement.

\begin{definition}[Refinement]\label{def:refinement}
  A target function \targetFunc \emph{refines} a source function \sourceFunc,
  denoted \refine{\sourceFunc}{\targetFunc}, if for every input:
  \begin{itemize}[leftmargin=*, topsep=0pt, partopsep=0pt]
    \item if \sourceFunc triggers immediate \ub, \targetFunc may exhibit
    any behavior~\cite{llvm_langref_immediate_UB};
    \item if \sourceFunc returns poison, \targetFunc may return poison or
    any other value, but must not trigger immediate
    \ub~\cite{llvm_langref_poison};
    \item if \sourceFunc returns a well-defined value, \targetFunc must
    return the same value and must not trigger immediate
    \ub~\cite{llvm_langref_well_defined}.
  \end{itemize}
\end{definition}
\noindent In the absence of \ub, this relation degenerates to equivalence.

\subsection{\llvm Translation Validation}
\label{subsec:translation-validation}
Translation validation checks each transformation result independently.
Given source and target functions \sourceFunc and \targetFunc,
the validator determines whether
\refine{\sourceFunc}{\targetFunc}.
A successful check verifies the transformation;
otherwise, it reports a counterexample showing that
\targetFunc does not refine \sourceFunc.

\subsubsection{Bounded \smt-Based Translation Validation}\leavevmode\\
Bounded \smt-based translation validation encodes the semantics
of the source and target functions as logical constraints over bounded domains,
such as bit-vectors with fixed-bitwidth and finitely unrolled loops. An \smt solver then searches
for an input on which \targetFunc does not refine \sourceFunc; if no such
input exists, the transformation is verified within
the encoded bounds~\cite{lee2021smtllvm,alivetwo}.
\alivetwo, the state-of-the-art \llvm translation
validator~\cite{alivetwo,llvm_instcombine_contributor_guide}, follows
this approach.

\myparagraph{Inherent Limitations of \alivetwo}
Despite its widespread adoption in the \llvm community,
\alivetwo inherits fundamental limitations from its underlying
bounded \smt-based design, particularly concerning
solver scalability, fixed-bitwidth reasoning, and bounded
loop unrolling.

\itemParagraph{Limitation: Solver Scalability}
\alivetwo suffers from scalability bottlenecks due to its
    whole-function formulation of refinement checking and its reliance
    on bit-blasting in the underlying \smt solver. It encodes
    complete source and target functions without decomposition,
    causing solver complexity to grow rapidly with
    instruction count and bitwidth~\cite{smtbitblasting}.
    Furthermore, bit-blasting lowers fixed-width arithmetic to
    Boolean circuits, which can become particularly expensive
    for transformations involving nonlinear bit-vector operations
    (\eg \texttt{bvudiv}) or generalized rewrites~\cite{lpg,hydra}.

\itemParagraph{Limitation: Fixed-Bitwidth Reasoning}
Since standard \smt bit-vector logics, such as the quantifier-free
    logic over fixed-size bit-vectors (QF\_BV), reason
    about fixed bitwidths~\cite{BarFT-RR-25}, \alivetwo~\cite{alivetwo}
    cannot validate transformations over symbolic bitwidths.
    Prior work therefore commonly bounds validation to selected
    bitwidths (\eg $64$ bits)~\cite{aliveone,hydra}.
    Consequently, validation at selected bitwidths
    leaves every other width unverified.
    Moreover, transformations with multiple bitwidths
    require validating each width combination separately,
    substantially increasing the validation overhead.

\itemParagraph{Limitation: Bounded Loop Unrolling}
\alivetwo handles transformations involving loops by unrolling
    them to a fixed depth~\cite{alivetwo}.
    Choosing this depth poses a dilemma: shallow unrolling misses
    bugs that manifest only at later iterations, while deeper bounds
    make the generated queries intractable.
    Consequently, bounded loop validation guarantees
    correctness only within the selected bound, leaving
    subsequent iterations unchecked.

\subsubsection{\itp-Based Translation Validation}

Interactive theorem provers (\itps) such as \coq~\cite{coq}
and \lean~\cite{lean,leanfour} provide a rigorous alternative
to \smt-based translation validation.
In an \itp, the semantics of the compiler \ir and
the target refinement relation are formalized in
higher-order logic, with correctness proofs
mechanically verified by a small, trusted kernel.
For example,
\vellvm~\cite{zhao2012vellvm,zhao2013ssa,zakowski2021vellvm,beck2024memory} formalizes
\llvm \ir semantics in \coq, while \leanmlir~\cite{leanmlir}
provides a \lean framework to validate transformations
across multiple \mlir dialects, including \llvm.

While \itp-based validation naturally handles
symbolic bitwidths and unbounded loops,
it fundamentally relies on manual proof construction---a
labor-intensive process requiring specialized formal methods expertise.
Recent advances in applying Large Language Models
(\llms) to automated theorem proving offer a
viable path to mitigate this burden.
In this work, we combine the foundational
correctness guarantees of \itps with
the automation of \llms to synthesize formal proofs for
\llvm transformations.

%% file: methodology.tex
This section presents the design of \proj,
a framework that integrates an \llm with \lean to
automate translation validation for \llvm transformations.

\begin{figure}[t]
  \centering

  \includegraphics[width=\linewidth]{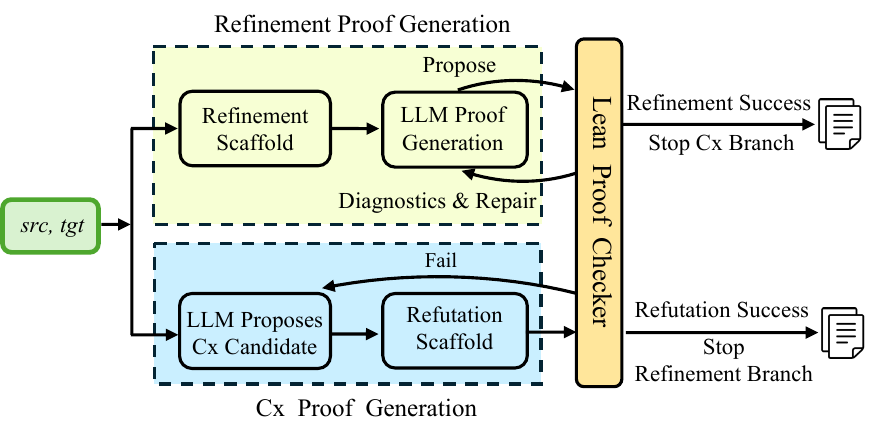}

  \caption{The workflow of \proj.
  \cx denotes counterexample.
  }
  \label{fig:workflow}
\end{figure}

\begin{figure*}[t]
  \centering

  \includegraphics[width=\textwidth]{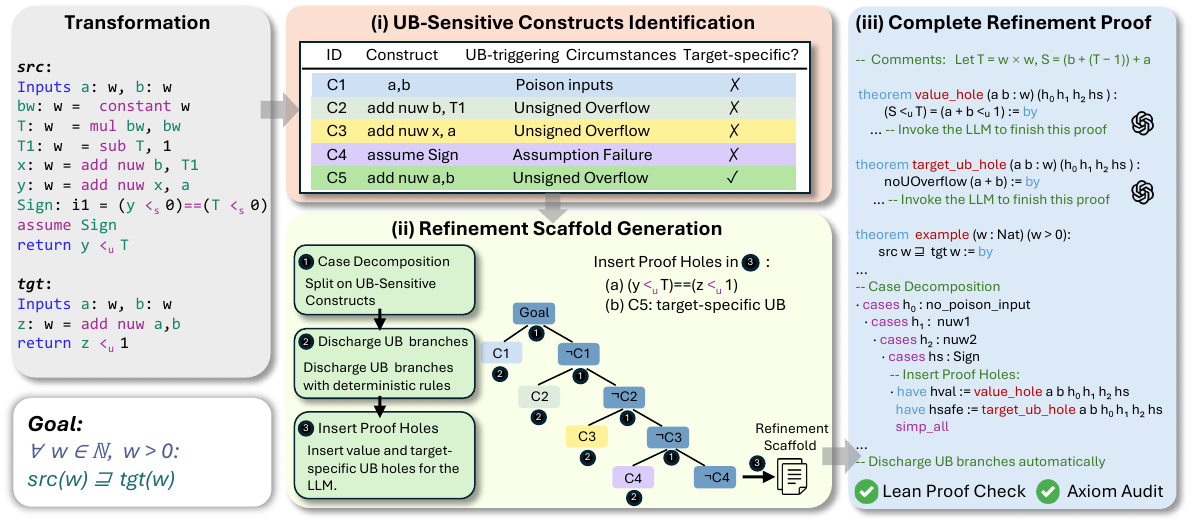}

  \caption{An illustrative example of refinement proof generation in \proj.
  }
  \label{fig:ie}
\end{figure*}

\Cref{fig:workflow} shows \proj's workflow.
For each transformation, \proj launches two branches \emph{in parallel},
each producing a proof for the \lean kernel to check:
\begin{itemize}[leftmargin=*, topsep=0pt, partopsep=0pt]
\item \textbf{Refinement branch.}
\proj first emits a refinement scaffold that discharges routine
obligations and leaves typed holes for the transformation-specific
proof obligations; the \llm later fills the holes.
\item \textbf{Counterexample branch.}
The \llm first proposes a counterexample candidate; \proj then emits
a refutation scaffold that turns the candidate into a
complete \lean proof of non-refinement.
\end{itemize}
\lean checks each proof, with its diagnostics guiding the \llm
to revise the refinement proof or propose a new counterexample.
Each branch repeats generation and checking until a proof is accepted
or the time budget expires. The first accepted proof establishes
validity or invalidity; \proj stops the other branch and
reports the result. If neither succeeds, \proj leaves the transformation
unresolved.

This section is organized as follows.
\Cref{subsec:refinement-generation,subsec:counterexample-generation}
present the refinement and counterexample branches for
loop-free, single-block transformations.
\Cref{subsec:loops} extends these workflows to a restricted
class of loop-containing transformations.
\Cref{subsec:implementation} describes the implementation and limitations.

\subsection{Refinement Proof Generation}
\label{subsec:refinement-generation}

\input{refinement}

\subsection{Counterexample Proof Generation}
\label{subsec:counterexample-generation}
\input{counterexample}

\subsection{Loop-Containing Transformations}
\label{subsec:loops}
\input{loops_preliminary}

\subsection{Implementation and Limitations}
\label{subsec:implementation}

\input{implementation}

%% file: refinement.tex
Generating complete refinement proofs with an \llm is costly,
particularly across diverse \ub cases.
To address this, we design a scaffold construction strategy that
delegates only transformation-specific reasoning to the \llm:
it decomposes the refinement goal (\refine{\sourceFunc}{\targetFunc})
via \ub-triggering conditions,
automatically discharges deterministic obligations,
and leaves only typed proof holes
for the \llm to complete.

To guide scaffold construction, we examine \leanmlir's \llvm dialect~\cite{leanmlir}
and identify \GuardRuleCount cases in which an operator or
instruction flag may produce \ub.
We call such operators and flags \emph{\ub-sensitive constructs}
and associate each with its precise \emph{\ub-triggering circumstance},
the condition under which each construct triggers \ub,
such as signed overflow for \texttt{nsw}
or a zero divisor for \texttt{udiv}.

\proj constructs the scaffold by splitting cases on these conditions
and automatically discharging proof obligations whenever the source yields \ub.
For well-defined source cases, \proj generates typed proof holes---establishing
value equivalence and absence of target \ub---to be completed by the \llm under
the corresponding branch hypotheses.
To guarantee soundness, \proj ensures that the generated proof passes \lean
kernel type-checking without open goals, strictly preserves the original
definitions and theorem statement, and relies exclusively on standard axioms~\cite{standard_axioms}.

\subsubsection{Illustrative Example}
\label{subsec:refinement-illustrative-example}
\Cref{fig:ie} illustrates refinement proof generation for a
real-world \llvm transformation~\cite{refinement_illustrative_issue}.
Let \(w\) be a positive bitwidth, and \(a\) and \(b\) be
\(w\)-bit inputs.
The transformation is:
\[
  \begin{gathered}
    \textcolor[HTML]{4F5D95}{\compactUnderbrace{
      (y<_{\mathrm{s}}0)=(T<_{\mathrm{s}}0)\;\vDash\;
      y<_{\mathrm{u}}T}{\sourceFunc(w)}}
    \;\Longrightarrow\;
    \textcolor[HTML]{1B6B78}{\compactUnderbrace{a+_{\texttt{nuw}}b<_{\mathrm{u}}1}{\targetFunc(w)}},\\
    \text{where}\;
    T=w \times w,\;
    y=(b+_{\texttt{nuw}}(T-1))+_{\texttt{nuw}}a.
  \end{gathered}
\]
Here, all arithmetic uses \(w\)-bit values. Multiplication and subtraction
wrap modulo \(2^w\), while \texttt{nuw} additions produce poison on
unsigned overflow. The operators \(<_{\mathrm{s}}\) and
\(<_{\mathrm{u}}\) denote signed and unsigned comparisons, respectively.
\(\sourceFunc(w)\) and \(\targetFunc(w)\) denote the source and target
functions instantiated with input bitwidth \(w\).
The refinement goal for the functions \sourceFunc and \targetFunc is:
\[
  \textcolor[HTML]{4F5D95}{\forall w\in\mathbb{N},\; w>0:}\quad
  \textcolor[HTML]{1B6B78}{\sourceFunc(w) \sqsupseteq \targetFunc(w).}
\]

As shown in \Cref{fig:ie}~(i),
\proj identifies five \ub-sensitive constructs:
poison inputs (C1), unsigned overflow in two source additions (C2, C3),
an assumption violation (C4), and unsigned overflow in the target addition (C5).
While C1--C4 stem from the source, C5 is target-specific.
Because C5 could cause the target to evaluate to poison for a well-defined source execution
(violating refinement), it must be proven unreachable and cannot be discharged automatically.

In \Cref{fig:ie}~(ii), \proj performs case analysis on C1--C4
to cover all source outcomes.
If any of C1--C3 holds, the source evaluates to poison, and a pre-existing
lemma discharges the obligation since the target produces no immediate \ub.
If C4 holds, the source triggers immediate \ub, vacuously satisfying refinement.

Otherwise, the source is well-defined.
\proj then generates two proof holes: a value equivalence hole
\((y <_{\mathrm{u}} T) = (z <_{\mathrm{u}} 1)\),
and a target-UB hole ensuring that \(a + b\) does not overflow (C5).
In \Cref{fig:ie}~(iii), the \llm fills these two holes under
the accumulated branch hypotheses (which rule out C1--C4).
Finally, the \lean kernel type-checks the assembled proof within the scaffold.

\subsubsection{Proof Scaffold Generation}
\label{subsec:refinement-scaffold-generation}

\proj first identifies \ub-sensitive constructs across both functions and
systematically splits on three categories:
\circled{1}~the source's \ub-triggering circumstances, ordered by when values are first demanded;
\circled{2}~target-specific circumstances that can trigger immediate \ub; and
\circled{3}~value-selection conditions (e.g., in \texttt{select} instructions~\cite{llvm_langref_select}),
which guides the \llm to prove each branch separately.
These splits produce proof obligations under distinct branch hypotheses, such that
the original refinement goal holds if and only if
all resulting obligations are discharged.

Each obligation is handled according to the source outcome.
For immediate \ub, \proj automatically discharges the obligation
because any target behavior is permitted~\cite{llvm_ub_refinement_lattice}.
For poison, it discharges the obligation once target immediate \ub is ruled out,
as the target may yield either poison or a well-defined value~\cite{llvm_langref_poison_replacement}.
For a well-defined value, \proj leaves one hole for value equality and
separate holes to rule out each target-specific \ub-triggering circumstance.
Finally, \proj validates the scaffold's syntactic and type correctness using \lean.

\subsubsection{LLM-Driven Proof Completion}
\label{subsec:refinement-llm-completion}

\proj asks the \llm to fill the remaining typed holes with proofs of value equivalence
and the absence of target-specific \ub.
A candidate proof is accepted when it satisfies the following conditions:
\begin{enumerate}[
  leftmargin=*,
]
  \item the completed proof passes kernel checking;
  \item the proof contains neither \texttt{sorry} nor \texttt{admit};
  \item the proof depends exclusively on the standard
        axioms~\cite{standard_axioms}; and
  \item the source and target definitions and the theorem
        statement remain unchanged.
\end{enumerate}

\myparagraph{Prompt Design}
The prompt template comprises three core elements.

\begin{itemize}[leftmargin=*]
  \item \textbf{System and Task.}
  The prompt assigns the \llm the role of a \lean theorem
  prover and identifies the target proof file. It instructs the \llm
  to finish the proof by replacing the \texttt{sorry} placeholders
  with valid proofs.

  \item \textbf{Instructions.}
  The prompt instructs the \llm to inspect
  proof goals and retrieve diagnostics.
  The prompt also specifies that the \llm may add auxiliary lemmas and raise
  resource limits, such as \texttt{maxHeartbeats} and \texttt{maxRecDepth},
  when necessary.

  \item \textbf{Hard Rules.}
  To preserve proof integrity, the prompt explicitly forbids changes to
  the refinement theorem statement or the source and target definitions.
\end{itemize}

%% file: counterexample.tex
\begin{figure*}[t]
  \centering

  \includegraphics[width=\textwidth]{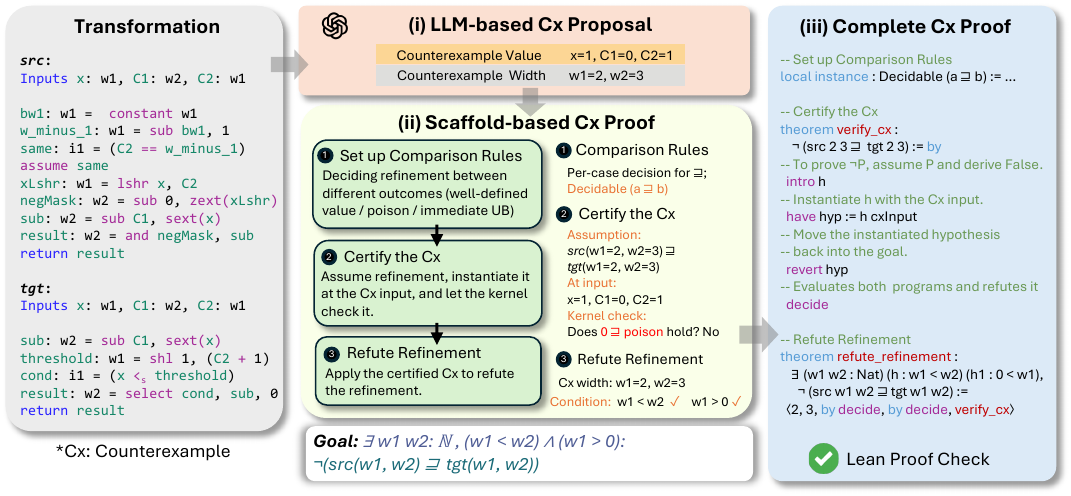}

  \caption{An illustrative example of counterexample proof generation in \proj.}
  \label{fig:counterexample-proof}
\end{figure*}

The counterexample branch refutes invalid transformations by turning a candidate
counterexample into a proof of non-refinement.
Because refinement quantifies universally over all admissible bitwidths and inputs,
a single counterexample suffices to disprove it.
The \llm is restricted solely to proposing a candidate counterexample with
concrete bitwidths and input values, after which a deterministic scaffold
constructs the complete \lean proof.
Because this scaffold introduces no additional axiomatic dependencies,
we omit the axiom audit from \Cref{subsec:refinement-llm-completion}.
The proof is accepted once it passes \lean kernel type-checking.

\subsubsection{Illustrative Example}
\label{subsec:counterexample-illustrative-example}

\Cref{fig:counterexample-proof} illustrates the counterexample proof
generation in \proj using an incorrect generalization of a real-world
\llvm transformation~\cite{hydra,illustrative_example2_issue}.
This transformation involves two symbolic bitwidths satisfying \(0 < w_1 < w_2\),
where inputs \(x\) and \(C_2\) are of width \(w_1\), whereas \(C_1\) is of width \(w_2\):
\begingroup
\[
  \begin{gathered}
    \textcolor[HTML]{4F5D95}{\compactUnderbrace{
      C_2=w_1-1 \;\vDash
      (0-\operatorname{zext}(x\gg_{\mathrm{u}}C_2))
      \mathbin{\&}(C_1-\operatorname{sext}(x))}{\sourceFunc(w_1,w_2)}}\\
    \mathrel{\textcolor[HTML]{1B6B78}{\Longrightarrow}}
    \textcolor[HTML]{1B6B78}{\compactUnderbrace{
      \bigl(x<_{\mathrm{s}}(1\ll(C_2+1))\bigr)
      \mathbin{?}(C_1-\operatorname{sext}(x))\mathbin{:}0}{\targetFunc(w_1,w_2)}.}
  \end{gathered}
\]
\endgroup
The source constructs a mask from the sign bit of \(x\), whereas the target
replaces this mask with a signed comparison and a conditional selection. The
counterexample goal is to establish that there exist admissible bitwidths for
which the source is not refined by the target:
\[
  \begin{aligned}
    &\textcolor[HTML]{4F5D95}{\exists w_1,w_2 \in \mathbb{N},\; 0 < w_1 < w_2:}\;
    \textcolor[HTML]{1B6B78}{\neg\bigl(\sourceFunc(w_1,w_2)
      \sqsupseteq \targetFunc(w_1,w_2)\bigr)}
  \end{aligned}
\]

The counterexample proof process comprises three stages.
First, the \llm proposes a candidate with bitwidths
\(w_1=2\) and \(w_2=3\) and input values \(x=1\), \(C_1=0\), and \(C_2=1\),
as shown in \Cref{fig:counterexample-proof}~(i).
Second, the scaffold constructs a formal proof of non-refinement from this candidate (\Cref{fig:counterexample-proof}~(ii)).
To enable proof by computation, the scaffold provides a computable \texttt{Decidable} instance
for \(a \sqsupseteq b\), covering well-defined values, poison, and immediate \ub.
This instance allows the \lean kernel to check that the concrete source and target outcomes
violate refinement, thereby certifying the counterexample.
To establish the contradiction, the scaffold assumes
\(\sourceFunc(2,3) \sqsupseteq \targetFunc(2,3)\), specializes this
assumption to the proposed counterexample inputs, and invokes kernel evaluation.
The source evaluates to a well-defined $0$, whereas the target evaluates to poison
under \llvm semantics~\cite{llvm_langref_poison} because \texttt{shl 1, 2} shifts a two-bit value
by its bitwidth.
The kernel therefore refutes \(0 \sqsupseteq \mathit{poison}\).
Together with proofs of the instantiated width constraints \(w_1<w_2\) (\ie, \(2<3\))
and \(0<w_1\) (\ie, \(0<2\)), this fixed-width contradiction refutes the refinement theorem.
Finally, \proj accepts the proof once it passes \lean kernel type-checking.

\subsubsection{LLM-Driven Counterexample Proposal}
\label{subsec:counterexample-llm-proposal}

\proj asks the \llm to propose a candidate counterexample with concrete bitwidths and input values.
If \proj fails to refute the refinement theorem using the candidate, it appends the \lean diagnostics
to the prompt and asks the \llm to generate a new candidate.
This process repeats until \lean verifies a proof of non-refinement or five attempts have been made.

\myparagraph{Prompt Design}
The prompt template follows the same structure as the refinement prompt in \Cref{subsec:refinement-llm-completion}.
\begin{itemize}[leftmargin=*]
  \item \textbf{System and Task.}
  The prompt assigns the \llm the role of a counterexample generator. It asks
  the \llm to find a counterexample that violates the
  refinement claim.

  \item \textbf{Instructions.}
  The prompt instructs the \llm to propose a counterexample by assigning
  concrete values to the symbolic bitwidths and inputs. If a candidate
  cannot be certified, the next prompt includes the Lean diagnostics and asks
  the \llm to propose a new counterexample candidate.

  \item \textbf{Hard Rules.}
  The \llm must save one candidate to the designated file as JSON,
  with
  complete \texttt{bitwidths} and \texttt{inputs} objects and no extra text.
  Before returning, it must check all constraints and confirm that the
  instantiated target function does not refine its source function.

\end{itemize}

\subsubsection{Scaffold-Based Proof Construction}
\label{subsec:counterexample-scaffold-proof}

Given a candidate counterexample proposed by the \llm,
\proj deterministically synthesizes a proof of non-refinement in two steps:

\myparagraph{Contradiction via Decidable Evaluation}
The scaffold assumes the claimed refinement theorem for contradiction and
specializes it with the proposed bitwidths and input values.
To evaluate the resulting concrete proposition, \proj leverages a \texttt{Decidable}
instance of the refinement relation ($a \sqsupseteq b$) that formalizes comparisons
across well-defined values, poison, and immediate \ub.
Kernel evaluation reduces the source and target expressions to concrete semantic outcomes;
because these outcomes violate refinement (\eg, deriving $0 \sqsupseteq \mathit{poison}$),
the decision procedure reduces the proposition to \texttt{False}.

\myparagraph{Precondition Discharge and Verification}
The scaffold then discharges the concrete bitwidth constraints (\eg, $0 < w_1 < w_2$)
via computation, assembling the counterexample and the contradiction into a complete proof
of the negated refinement theorem.
The final proof contains no remaining holes and is validated by \lean kernel type-checking.

%% file: loops_preliminary.tex
For loop-free transformations, refinement can be established by
directly comparing the source and target behaviors for all
possible inputs. For transformations involving loops, however,
the proof must also account for all feasible loop iteration counts, since
a loop may execute an arbitrary number of iterations depending
on the inputs. To address this challenge, \alivetwo applies
bounded loop unrolling, checking refinement only up
to a selected unrolling bound and thereby potentially missing
counterexamples that require more iterations~\cite{alivetwo}.
By contrast, \proj establishes refinement by induction, yielding
a single correctness proof that covers all feasible loop iteration counts
without relying on a fixed unrolling bound.

In this paper, \proj provides preliminary support for a restricted
class of loop-containing transformations, from a source
function containing \llvm's canonical loop
form~\cite{llvm_loop_terminology,llvm_loop_pass_manager} to a
loop-free target function with a single basic block. Specifically, the source function
consists of three blocks: an entry preheader, a single
self-looping body block with one back edge,
and a dedicated exit block that returns immediately.
This fixed topology enables \proj to reuse a fixed proof
scaffold and is grounded in practice:
canonical loop form is enforced before the loop-pass execution,
and the single-block, single-back-edge form is targeted by several
loop idiom recognizers~\cite{llvm_loop_pass_manager,llvm_loop_idiom_recognize}.

For general loop-containing transformations, however, the
proof scaffold varies with the source function's control-flow
graph (CFG). This variation arises because inductive
proofs require every cycle in the CFG to pass through
an invariant-bearing cutpoint~\cite{floyd1993assigning},
resulting in CFG-specific invariants and induction obligations.
Accordingly, we leave support for general loop structures to future work.

\subsubsection{Loop-Aware Refinement Scaffold and Proof Generation}

Following the loop-free workflow in \Cref{subsec:refinement-generation},
\proj first constructs a loop-aware scaffold by extending the
loop-free scaffold in \Cref{subsec:refinement-scaffold-generation}
to loop-containing transformations and then invokes the \llm
to fill the proof holes and complete the proof.

For loop-containing transformations, \proj reuses the loop-free scaffold's
case decomposition, refinement rules, and target-specific \ub holes.
In the source entry block, it splits cases and automatically applies the
refinement rules to discharge obligations for poison and immediate \ub outcomes.
For the target function, \proj generates
target-specific \ub holes to rule out \ub in the target
when the source returns a well-defined value.
For well-defined source outcomes, \proj's loop-free scaffold uses
a single value hole requiring the source and target functions
to return the same value.
In contrast, for transformations involving loops, \proj replaces this
single hole with loop-specific holes that jointly specify the
proof obligations for refinement over an arbitrary number of iterations.
\Cref{tab:loops-hole-taxonomy} provides an overview of the
general and loop-specific holes in the refinement scaffold.
Finally, \proj invokes the \llm to provide the loop invariant and
fill the proof holes in the scaffold and checks the completed
proof following the same procedure in \Cref{subsec:refinement-llm-completion}.

\begin{figure}[t]
  \centering
  \includegraphics[width=\linewidth]{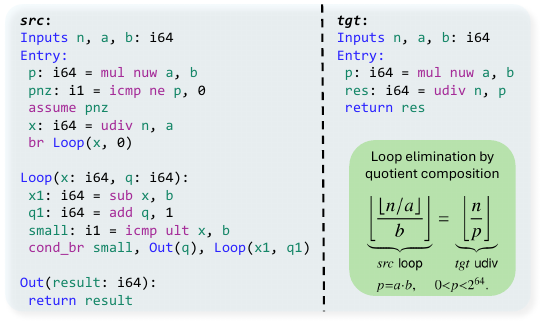}
  \caption{An illustrative example for a loop-containing transformation}
  \label{fig:loop-elimination-example}
\end{figure}

\myparagraph{Illustrative Example}
\Cref{fig:loop-elimination-example} shows an illustrative loop-containing
transformation that replaces a quotient computation by repeated subtraction
with an unsigned division. The source loop implements Euclidean
division~\cite{boute1992euclidean}: for $x,b\in\mathbb{N}$ with $b>0$, the
unique quotient--remainder pair $(q,r)$ satisfies $x=q\cdot b+r$ and
$0\leq r<b$, where $q=\lfloor x/b\rfloor$ counts the subtractions. Starting
from \texttt{x = udiv n, a} and $q=0$, the source repeatedly updates $(x,q)$ to
$(x-b,q+1)$ while $x\geq b$ and then returns $q$. It thus computes
$\lfloor\lfloor n/a\rfloor/b\rfloor=\lfloor n/(a\cdot b)\rfloor$, which the
target computes directly using \texttt{udiv n, p}, where $p=a\cdot b$.

\myparagraph{Refinement Proof for the Example}
\proj's refinement proof scaffold first applies the
case decomposition process to discharge the source entry block's poison and
immediate \ub cases (\eg $p=0$ violates \texttt{assume pnz}).
For the remaining cases, the scaffold leaves target-specific \ub
obligations and three loop-specific holes:
\begin{enumerate}[
  label=\arabic*.,
  leftmargin=*,
  align=left,
  labelsep=0.5em
]
  \item \emph{Target-specific \ub.}
        Proofs that the target's \texttt{mul nuw} cannot overflow and
        that $p\neq 0$ for \texttt{udiv n, p}.
  \item \emph{Loop invariant.}
        A predicate $I(n,a,b,x,q)$ relating the \targetFunc
        inputs ($n$, $a$, and $b$) to the \sourceFunc loop
        state ($x$, and $q$).
  \item \emph{Inductive step.}
        A proof that $I$ is preserved by $(x,q)\mapsto(x-b,q+1)$ when
        $x\geq b$ and implies $q=\lfloor n/p\rfloor$ when $x<b$.
  \item \emph{Final assembly.}
        A proof establishing $I$ at loop entry, where
        $(x,q)=(\lfloor n/a\rfloor,0)$, and using the inductive step
        to show that the source loop's result equals $\lfloor n/p\rfloor$.
\end{enumerate}

\noindent The \llm completes the proof by filling these holes as follows:
\begin{enumerate}[
  label=\arabic*.,
  leftmargin=*,
  align=left,
  labelsep=0.5em
]
  \item \emph{Target-specific \ub.}
        It uses $a\cdot b<2^{64}$ and $p\neq 0$ from the source entry
        conditions to exclude overflow in \texttt{mul nuw} and division
        by zero in \texttt{udiv n, p}.
  \item \emph{Loop invariant.}
        It defines $I$ to require well-defined inputs and loop values,
        $0<a\cdot b<2^{64}$, and
        $q+\lfloor x/b\rfloor=\lfloor n/(a\cdot b)\rfloor$.
        Here, $q$ and $\lfloor x/b\rfloor$ count completed and remaining
        subtractions, respectively.
  \item \emph{Inductive step.}
        On the back edge ($x\geq b$), it derives $q+1<2^{64}$ from $I$,
        so neither $x-b$ nor $q+1$ wraps. It then uses
        $\lfloor x/b\rfloor=1+\lfloor(x-b)/b\rfloor$ to prove that
        $(x,q)\mapsto(x-b,q+1)$ preserves $I$.
        At exit ($x<b$), $\lfloor x/b\rfloor=0$ gives
        $q=\lfloor n/p\rfloor$, matching the target result.
  \item \emph{Final assembly.}
        At loop entry, $x=\lfloor n/a\rfloor$ and $q=0$.
        The source entry conditions and the identity
        $\lfloor\lfloor n/a\rfloor/b\rfloor=\lfloor n/(a\cdot b)\rfloor$
        establish $I$ for this initial state. The \llm combines this
        initialization proof with the inductive step to prove refinement
        for any number of loop iterations.
\end{enumerate}
\proj checks the completed proof following
\Cref{subsec:refinement-llm-completion}.

\begin{table}[t]
  \centering
  \caption{General and loop-specific holes in the preliminary
  loop-containing scaffold.}
  \label{tab:loops-hole-taxonomy}
  \footnotesize
  \renewcommand{\arraystretch}{1.08}
  \setlength{\tabcolsep}{3pt}
  \begin{tabular}{@{}
    >{\centering\arraybackslash}p{0.15\columnwidth}
    >{\raggedright\arraybackslash}p{0.80\columnwidth}@{}}
    \toprule
    \textbf{Type} & \textbf{Hole and Role} \\
    \midrule
    General &
      \hangindent=1.2em\hangafter=1
      \textbullet\enspace \emph{Target-specific \ub}: rules out each relevant
      target-side \ub condition whenever the source is defined. \\
    \midrule
    \multirow{3}{*}{\begin{tabular}{@{}c@{}}Loop-\\specific\end{tabular}} &
      \hangindent=1.2em\hangafter=1
      \textbullet\enspace \emph{Loop invariant}: summarizes the reachable
      source-loop state and function inputs. \\
    \cmidrule(l){2-2}
    & \hangindent=1.2em\hangafter=1
      \textbullet\enspace \emph{Inductive step}: preserves the invariant and
      proves equality with the target result at loop exit. \\
    \cmidrule(l){2-2}
    & \hangindent=1.2em\hangafter=1
      \textbullet\enspace \emph{Final assembly}: establishes the invariant
      at loop entry, then uses the inductive proof to establish refinement
      for any number of loop iterations. \\
    \bottomrule
  \end{tabular}
\end{table}

\subsubsection{Loop-Aware Counterexample Certification}

The counterexample branch extends the loop-free workflow in
\Cref{subsec:counterexample-generation}.
The \llm proposes a counterexample candidate with concrete bitwidths,
inputs, and an execution bound
that limits the loop iteration count and ensures evaluation terminates.
The scaffold accepts a counterexample candidate only if the loop
execution completes within this bound.
Otherwise, the scaffold rejects the candidate and requests a new one.
Given this candidate, \proj reuses the deterministic
stages from \Cref{subsec:counterexample-scaffold-proof}:
decidable comparison of execution outcomes, certification of the concrete
candidate, and construction of a proof of non-refinement from the certified counterexample.
Following the loop-free workflow in \Cref{subsec:counterexample-generation},
\proj allows at most five counterexample generation attempts per transformation.

%% file: implementation.tex
We build \proj on top of the \llvm dialect in \leanmlir~\cite{leanmlir},
a \lean~4 framework that formalizes the semantics of \mlir programs:
it parses \mlir code into well-typed \lean terms with precise
per-operation semantics, so that properties of \mlir programs become
\lean theorems.
Using \mlir's translation mechanisms~\cite{mlir_llvm_ir_target},
\llvm \ir converts bidirectionally to and from the \llvm dialect
without loss of semantic information~\cite{mlir_llvm_dialect}.

However, \leanmlir originally modeled only a simplified subset of the
\llvm dialect. To enable translation validation of real-world
transformations, we extended it along three dimensions:
\circled{1}~refining its semantic foundations to support immediate \ub
and transformation preconditions,
\circled{2}~adding missing bit-manipulation operators (\eg \texttt{ctpop},
\texttt{cttz}, \texttt{ctlz}), and
\circled{3}~generalizing the framework to multiple symbolic-bitwidth
parameters and multi-block transformations.

In \proj, the \llm operates as an agent: we use the Codex
CLI~\cite{codex_cli} with the \texttt{gpt-5.5}
model~\cite{openai_gpt55_model} to perform proof completion in the
refinement branch and counterexample proposal in the counterexample
branch.
\proj itself remains a deterministic harness, accepting the agent's
output only after the \lean kernel checks the resulting proof.

Our formalization currently targets integer instructions, and our
workflows support loop-free, single-block transformations plus a
restricted class of loop-containing ones. Extending the formalization
to memory operations, floating-point arithmetic, and richer loop
classes is left as future work.

%% file: evaluation.tex
This section presents our extensive evaluation of \proj,
demonstrating its effectiveness in automating translation validation for
\llvm transformations. We organize our evaluation around
four research questions:

\begin{itemize}[leftmargin=*]
    \item \textbf{\rqone:} \rqoneContent
    \item \textbf{\rqtwo:} \rqtwoContent
    \item \textbf{\rqthree:} \rqthreeContent
    \item \textbf{\rqfour:} \rqfourContent
\end{itemize}

We intentionally separate loop-free (\rqone and \rqtwo) and loop-containing
transformations (\rqthree) because \alivetwo offers different guarantees.
For loop-free transformations, \alivetwo can validate them under its modeled semantics,
while for loop-containing transformations, it only provides bounded
guarantees due to its finite loop unrolling~\cite{alivetwo}.

\myparagraph{Setup}
All experiments were conducted on an Ubuntu 22.04 (64-bit)
server equipped with a 32-core Intel Xeon Gold 5217 CPU @ 3.00GHz
and 376 GB of RAM. We applied a time limit of \exptime per
transformation across all RQs.
Since \proj does not support \texttt{undef} for the reasons discussed in
\Cref{subsec:llvm-transformations},
we ran \alivetwo with
\texttt{--disable-undef-input}; this configuration ensures
a fair comparison because our dataset has no other mechanism
for introducing \texttt{undef}.
As the underlying LLM,
\proj uses \texttt{gpt-5.5} with extra-high reasoning effort.

\subsection{Dataset}
Our dataset comprises \ValBenchCount real-world \llvm transformation
instances from the \llvm GitHub repository.
\Cref{tab:dataset-composition} summarizes their distribution by loop structure,
validity, bitwidth, and research question.

\begin{table}[t]
  \centering
  \caption{Dataset Composition}
  \label{tab:dataset-composition}
  \footnotesize
  \renewcommand{\arraystretch}{1.0}
  \setlength{\tabcolsep}{4pt}
  \begin{tabular}{@{}lllrrr@{}}
    \toprule
    & & & \multicolumn{2}{c}{\textbf{Bitwidth}} & \\
    \cmidrule(lr){4-5}
    \textbf{Loop Structure} & \textbf{RQ} & \textbf{Validity}
      & \textbf{Fixed} & \textbf{Symbolic} & \textbf{Total} \\
    \midrule
    \multirow{2}{*}{Loop-free}
      & \rqone & Valid   & \DsFix & \DsSym & \DsTotal \\
      & \rqtwo & Invalid & \DsMutFix & \DsMutSym & \DsMut \\
    \midrule
    \multirow{2}{*}{Loop-containing}
      & \rqthree & Valid   & \RqThreeValidCaseCountFixed
                            & \RqThreeValidCaseCountSymbolic
                            & \RqThreeValidCaseCount \\
      & \rqthree & Invalid & \RqThreeInvalidCaseCountFixed
                            & \RqThreeInvalidCaseCountSymbolic
                            & \RqThreeInvalidCaseCount \\
    \midrule
    Overall & -- & -- & \ValBenchCountFixed & \ValBenchCountSymbolic
                     & \ValBenchCount \\
    \bottomrule
  \end{tabular}
\end{table}

\myparagraph{Loop-Free Dataset}
The loop-free dataset contains fixed- and symbolic-bitwidth cases.

In the
fixed-bitwidth portion, \alivetwo proves the \DsFs valid cases and refutes the
\DsMutFs invalid cases within its standard 10\,s time limit.
The remaining
\DsFt valid cases and \DsMutFt invalid cases involve nonlinear bit-vector
operations, interdependent preconditions, and relatively large but common
concrete bitwidths (\eg \texttt{i32} and \texttt{i64}).  These structures make it
challenging for \smt-based methods such as \alivetwo
to finish within the standard 10\,s time limit~\cite{smtbitblasting}.
We include these cases to assess whether \proj can validate or refute
transformations that are challenging for \smt-based methods.

In the symbolic-bitwidth portion, we derive each valid case
by applying the transformation-generalization techniques
developed in prior work~\cite{hydra,lpg} to its fixed-bitwidth
counterpart.  We manually construct each invalid case from
its fixed-bitwidth counterpart by replacing the concrete
integer bitwidth with a symbolic one, because these techniques
do not support the generalization of invalid transformations.

\myparagraph{Loop-Containing Dataset}
The loop-containing dataset comprises \RqThreeValidCaseCount valid and
\RqThreeInvalidCaseCount invalid transformations.
Each group contains equal numbers of fixed- and symbolic-bitwidth cases.
Each transformation replaces a three-basic-block source function
with a loop-free target function with a single basic block.
We derive each symbolic-bitwidth case from its fixed-bitwidth counterpart
by manually replacing the concrete integer bitwidth with a symbolic one
because prior transformation-generalization techniques~\cite{hydra,lpg}
do not support loop-containing transformations.

\subsection{\rqone: Validating Transformations}
\label{sec:rq1}

\begin{table}[t]
  \centering
  \caption{\proj versus \alivetwo on \RqOneCount valid transformations.}
  \label{tab:rq1}
  \footnotesize
  \renewcommand{\arraystretch}{1.0}
  \setlength{\tabcolsep}{6pt}
  \begin{tabular}{@{}lrrrrr@{}}
    \toprule
    & & \multicolumn{2}{c}{\textbf{\proj}} & \multicolumn{2}{c}{\textbf{\alivetwo}} \\
    \cmidrule(lr){3-4}\cmidrule(lr){5-6}
    \textbf{Stratum} & \textbf{$N$} & \textbf{Proved} & \textbf{Mean\,(s)}
                                    & \textbf{Proved} & \textbf{Mean\,(s)} \\
    \midrule
    Symbolic bitwidth & \RqOneSymCount & \RqOneSymVerified & \RqOneSymMean & n/a & n/a \\
    Fixed bitwidth    & \RqOneFixCount & \RqOneFixVerified & \RqOneFixMean
                      & \AliveFixedBWVerifiedCount & \AliveFixedBWMean \\
    \midrule
    Total             & \RqOneCount & \RqOneVerified & \RqOneMean
                      & \AliveFixedBWVerifiedCount & \AliveFixedBWMean \\
    \bottomrule
  \end{tabular}
\end{table}

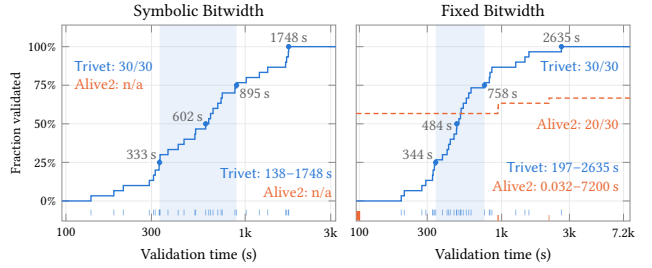
\begin{figure}[t]
  \centering
  \resizebox{\columnwidth}{!}{%
  \begin{tikzpicture}
    \begin{axis}[
      width=0.66\columnwidth, height=4.0cm, scale only axis,
      xmode=log, xmin=95, xmax=3200, ymin=-0.13, ymax=1.14,
      title={\normalsize Symbolic Bitwidth},
      title style={yshift=-1.4ex, text height=1.8ex, text depth=0.4ex},
      xlabel={Validation time (s)}, ylabel={Fraction validated},
      xtick={100,300,1000,3000}, xticklabels={100,300,1k,3k},
      ytick={0,0.25,0.5,0.75,1},
      yticklabels={0\%,25\%,50\%,75\%,100\%},
      grid=major, grid style={draw=black!8},
      axis line style={draw=black!45}, tick style={draw=black!45},
      label style={font=\small}, tick label style={font=\footnotesize},
      ylabel style={font=\footnotesize,
        at={(axis description cs:-0.13,0.5)}, anchor=south},
      clip mode=individual,
    ]
      \addplot[draw=none, fill=L3blue, fill opacity=0.10]
        coordinates {(333,-0.13) (895,-0.13) (895,1.14) (333,1.14)} \closedcycle;
      \addplot[no marks, thick, color=L3blue] coordinates {(95,0.0000) (138,0.0000) (138,0.0333) (185,0.0333) (185,0.0667) (209,0.0667) (209,0.1000) (292,0.1000) (292,0.1333) (308,0.1333) (308,0.1667) (315,0.1667) (315,0.2000) (331,0.2000) (331,0.2333) (333,0.2333) (333,0.2667) (335,0.2667) (335,0.3000) (371,0.3000) (371,0.3333) (423,0.3333) (423,0.3667) (456,0.3667) (456,0.4000) (524,0.4000) (524,0.4333) (529,0.4333) (529,0.4667) (602,0.4667) (602,0.5000) (626,0.5000) (626,0.5333) (641,0.5333) (641,0.5667) (662,0.5667) (662,0.6000) (703,0.6000) (703,0.6333) (734,0.6333) (734,0.6667) (745,0.6667) (745,0.7000) (876,0.7000) (876,0.7333) (895,0.7333) (895,0.7667) (1014,0.7667) (1014,0.8000) (1200,0.8000) (1200,0.8333) (1335,0.8333) (1335,0.8667) (1681,0.8667) (1681,0.9000) (1703,0.9000) (1703,0.9333) (1741,0.9333) (1741,0.9667) (1748,0.9667) (1748,1.0000) (3200,1.0000)};
      \addplot[only marks, mark=|, mark size=1.7pt, color=L3blue, opacity=0.5]
        coordinates {(138,-0.075) (185,-0.075) (209,-0.075) (292,-0.075) (308,-0.075) (315,-0.075) (331,-0.075) (333,-0.075) (335,-0.075) (371,-0.075) (423,-0.075) (456,-0.075) (524,-0.075) (529,-0.075) (602,-0.075) (626,-0.075) (641,-0.075) (662,-0.075) (703,-0.075) (734,-0.075) (745,-0.075) (876,-0.075) (895,-0.075) (1014,-0.075) (1200,-0.075) (1335,-0.075) (1681,-0.075) (1703,-0.075) (1741,-0.075) (1748,-0.075)};
      \addplot[only marks, mark=*, mark size=1.25pt, color=L3blue]
        coordinates {(333,0.25) (602,0.50) (895,0.75) (1748,1.00)};
      \node[font=\small,text=black!60,anchor=south east,inner sep=0.5pt,xshift=-1pt,yshift=1pt] at (axis cs:333,0.25) {333\,s};
      \node[font=\small,text=black!60,anchor=south east,inner sep=0.5pt,xshift=-1pt,yshift=1pt] at (axis cs:602,0.50) {602\,s};
      \node[font=\small,text=black!60,anchor=north west,inner sep=0.5pt,xshift=1pt,yshift=-1pt] at (axis cs:895,0.75) {895\,s};
      \node[font=\small,text=black!60,anchor=south,inner sep=0.5pt,yshift=2pt]
        at (axis cs:1748,1.00) {1748\,s};
      \node[font=\small,text=L3blue,anchor=west,inner sep=0.5pt]
        at (axis cs:110,0.87) {\proj: \RqOneSymVerified/\RqOneSymCount};
      \node[font=\small,text=L3orange,anchor=west,inner sep=0.5pt]
        at (axis cs:110,0.76) {\alivetwo: n/a};
      \node[font=\small,text=L3blue,anchor=east,inner sep=0.5pt]
        at (axis cs:3100,0.18)
        {\proj: 138--1748\,s};
      \node[font=\small,text=L3orange,anchor=east,inner sep=0.5pt]
        at (axis cs:3100,0.06) {\alivetwo: n/a};
    \end{axis}
  \end{tikzpicture}\hspace{-0.2em}
  \begin{tikzpicture}
    \begin{axis}[
      width=0.66\columnwidth, height=4.0cm, scale only axis,
      xmode=log, xmin=95, xmax=8000, ymin=-0.13, ymax=1.14,
      title={\normalsize Fixed Bitwidth},
      title style={yshift=-1.4ex, text height=1.8ex, text depth=0.4ex},
      xlabel={Validation time (s)},
      xtick={100,300,1000,3000,7200}, xticklabels={100,300,1k,3k,7.2k},
      ytick={0,0.25,0.5,0.75,1}, yticklabels={,,,,},
      grid=major, grid style={draw=black!8},
      axis line style={draw=black!45}, tick style={draw=black!45},
      label style={font=\small}, tick label style={font=\footnotesize},
      clip mode=individual,
    ]
      \addplot[draw=none, fill=L3blue, fill opacity=0.10]
        coordinates {(344,-0.13) (758,-0.13) (758,1.14) (344,1.14)} \closedcycle;
      \addplot[no marks, thick, color=L3blue] coordinates {(95,0.0000) (197,0.0000) (197,0.0333) (207,0.0333) (207,0.0667) (275,0.0667) (275,0.1000) (293,0.1000) (293,0.1333) (327,0.1333) (327,0.1667) (330,0.1667) (330,0.2000) (337,0.2000) (337,0.2333) (344,0.2333) (344,0.2667) (384,0.2667) (384,0.3000) (414,0.3000) (414,0.3333) (424,0.3333) (424,0.3667) (436,0.3667) (436,0.4000) (463,0.4000) (463,0.4333) (482,0.4333) (482,0.4667) (484,0.4667) (484,0.5000) (501,0.5000) (501,0.5333) (514,0.5333) (514,0.5667) (519,0.5667) (519,0.6000) (538,0.6000) (538,0.6333) (565,0.6333) (565,0.6667) (587,0.6667) (587,0.7000) (610,0.7000) (610,0.7333) (758,0.7333) (758,0.7667) (820,0.7667) (820,0.8000) (829,0.8000) (829,0.8333) (854,0.8333) (854,0.8667) (1257,0.8667) (1257,0.9000) (1459,0.9000) (1459,0.9333) (1557,0.9333) (1557,0.9667) (2635,0.9667) (2635,1.0000) (8000,1.0000)};
      \addplot[no marks, thick, densely dashed, color=L3orange]
        coordinates {(95,0.5667) (941.0,0.5667) (941.0,0.6000) (948.106,0.6000) (948.106,0.6333) (2155.0,0.6333) (2155.0,0.6667) (8000,0.6667)};
      \addplot[only marks, mark=|, mark size=1.7pt, color=L3blue, opacity=0.5]
        coordinates {(197,-0.075) (207,-0.075) (275,-0.075) (293,-0.075) (327,-0.075) (330,-0.075) (337,-0.075) (344,-0.075) (384,-0.075) (414,-0.075) (424,-0.075) (436,-0.075) (463,-0.075) (482,-0.075) (484,-0.075) (501,-0.075) (514,-0.075) (519,-0.075) (538,-0.075) (565,-0.075) (587,-0.075) (610,-0.075) (758,-0.075) (820,-0.075) (829,-0.075) (854,-0.075) (1257,-0.075) (1459,-0.075) (1557,-0.075) (2635,-0.075)};
      \draw[color=L3orange,line width=2.2pt,line cap=butt,xshift=1.2pt]
        (axis cs:95,-0.125) -- (axis cs:95,-0.065);
      \addplot[only marks, mark=|, mark size=1.7pt, color=L3orange, opacity=0.7]
        coordinates {(941,-0.110) (948.106,-0.110) (2155,-0.110)};
      \addplot[only marks, mark=*, mark size=1.25pt, color=L3blue]
        coordinates {(344,0.25) (484,0.50) (758,0.75) (2635,1.00)};
      \node[font=\small,text=black!60,anchor=south east,inner sep=0.5pt,xshift=-1pt,yshift=1pt] at (axis cs:344,0.25) {344\,s};
      \node[font=\small,text=black!60,anchor=east,inner sep=0.5pt,xshift=-2pt,yshift=-1pt] at (axis cs:484,0.50) {484\,s};
      \node[font=\small,text=black!60,anchor=north west,inner sep=0.5pt,xshift=1pt,yshift=-1pt] at (axis cs:758,0.75) {758\,s};
      \node[font=\small,text=black!60,anchor=south,inner sep=0.5pt,yshift=2pt]
        at (axis cs:2635,1.00) {2635\,s};
      \node[font=\small,text=L3blue,anchor=east,inner sep=0.5pt]
        at (axis cs:6800,0.87) {\proj: \RqOneFixVerified/\RqOneFixCount};
      \node[font=\small,text=L3orange,anchor=east,inner sep=0.5pt]
        at (axis cs:6800,0.50)
        {\alivetwo: \AliveFixedBWVerifiedCount/\AliveFixedBWCount};
      \node[font=\small,text=L3blue,anchor=east,inner sep=0.5pt]
        at (axis cs:6800,0.22) {\proj: 197--2635\,s};
      \node[font=\small,text=L3orange,anchor=east,inner sep=0.5pt]
        at (axis cs:6800,0.10) {\alivetwo: 0.032--7200\,s};
    \end{axis}
  \end{tikzpicture}
  }
  \caption{
    Validation time distributions for \rqone (log scale).
  }
  \label{fig:rq1-dist}
\end{figure}

In \rqone, we investigate whether \proj can establish refinement for
real-world \llvm transformations and compare its validation capability with
\alivetwo.
\Cref{tab:rq1} summarizes the aggregate results for both systems on the same
\RqOneCount transformations and \Cref{fig:rq1-dist} shows the
validation-time distributions.

\subsubsection{Experimental Results}

\proj validates all \RqOneCount transformations,
with median and mean validation times of \RqOneMedian{}\,s and
\RqOneMean{}\,s, respectively.  \alivetwo applies only to the
\AliveFixedBWCount fixed-bitwidth transformations and validates
\AliveFixedBWVerifiedCount; the remaining \AliveFixedBWTimeoutCount
runs do not complete within the time limit.

For the symbolic-bitwidth cases, \proj validates all \RqOneSymCount cases
in 138--1748\,s, with a median of \RqOneSymMedian{}\,s and a
mean of \RqOneSymMean{}\,s.  Each generated theorem is parametric in the
bitwidth and establishes refinement for every width satisfying the required
constraints (\eg \(\smash{w \bmod 2 = 0}\)).  Since \alivetwo does not support
transformations with symbolic bitwidths, \Cref{tab:rq1} reports
``n/a'' for these cases.

For the fixed-bitwidth cases, \alivetwo takes less than 1\,s
to validate each of the \DsFs cases,
whereas \proj validates the same cases
with a median of \RqOneFsMedian{}\,s and a mean of \RqOneFsMean{}\,s.
In the remaining \DsFt challenging cases, \alivetwo validates
only \AliveFtVerified within the time limit of \exptime; its successful runs have a
median of \AliveFtMedian{}\,s and a mean of \AliveFtMean{}\,s.
In contrast, \proj validates all \DsFt challenging cases
in 207--1257\,s, with a median of 492\,s.
Overall, \alivetwo is faster on the \DsFs cases but validates
only a subset of the \DsFt challenging cases, whereas \proj validates all
fixed-bitwidth cases.

\subsubsection{Result Analysis}

The symbolic-bitwidth results show that a single \proj proof
establishes refinement for all admissible widths rather than
for one concrete instantiation, providing a width-independent
guarantee outside \alivetwo's supported scope.  For fixed
bitwidths, \alivetwo exhibits a performance cliff
as proof difficulty increases. Similarly, the original \alivetwo
study found that extending the timeout from one to five minutes
yielded less than 5\% additional results~\cite{alivetwo}.
By contrast, \proj exhibits more consistent runtimes across
the fixed-bitwidth cases.

The performance cliff stems primarily from two factors:
\alivetwo's whole-function refinement formulation and the
bit-blasting in the \smt solver.
\alivetwo encodes the complete source and target functions
rather than decomposing validation into smaller independent
obligations, producing increasingly difficult queries as the
transformations grow more complex.
Moreover, this difficulty is exacerbated by the bit-blasting
mechanism, especially for transformations
involving nonlinear operators (\eg div, mul, and urem).
Although bit-blasting enables automated reasoning via SAT,
expanding \(w\)-bit operations into Boolean circuits obscures
their algebraic structure and makes particular arithmetic
operations expensive to solve~\cite{barth2026lazy,hadarean2014tale,smtbitblasting}.
In contrast, \proj leverages the \llm's algebraic reasoning capabilities
to decompose complex proof obligations, completing
the proof step by step, thereby avoiding this performance cliff.

Prior work shows that bit-blasting can be extremely expensive
at common bitwidths such as 32 bits~\cite{smtbitblasting};
whole-function refinement checking can further amplify this cost.
For example, consider a simple transformation
\(\bigl((x/_{\mathrm{u}}y)\mathbin{\cdot}y \le x\bigr) \Longrightarrow \AlgTrue\),
where \(/{_{\mathrm{u}}}\) denotes unsigned integer division.
Although this transformation is clearly valid under \llvm semantics,
\alivetwo times out on its \texttt{i32} instance after two hours,
whereas \proj validates the same \texttt{i32} instance in 339\,s.

\subsection{\rqtwo: Refuting Invalid Transformations}
\label{sec:rq2}

\begin{table}[t]
  \centering
  \caption{
    \proj versus \alivetwo on \RqTwoCount invalid transformations.
  }
  \label{tab:rq2}
  \footnotesize
  \renewcommand{\arraystretch}{1.0}
  \setlength{\tabcolsep}{6pt}
  \begin{tabular}{@{}lrrrrr@{}}
    \toprule
    & & \multicolumn{2}{c}{\textbf{\proj}} & \multicolumn{2}{c}{\textbf{\alivetwo}} \\
    \cmidrule(lr){3-4}\cmidrule(lr){5-6}
    \textbf{Stratum} & \textbf{$N$} & \textbf{Refuted} & \textbf{Mean\,(s)}
                                    & \textbf{Refuted} & \textbf{Mean\,(s)} \\
    \midrule
    Symbolic bitwidth & \RqTwoSymCount & \RqTwoSymRefuted & \RqTwoSymMean
                      & n/a & n/a \\
    Fixed bitwidth    & \RqTwoFixCount & \RqTwoFixRefuted & \RqTwoFixMean
                      & \AliveCexRefuted & \AliveCexMean \\
    \midrule
    Total             & \RqTwoCount & \RqTwoRefuted & \RqTwoMean
                      & \AliveCexRefuted & \AliveCexMean \\
    \bottomrule
  \end{tabular}
\end{table}

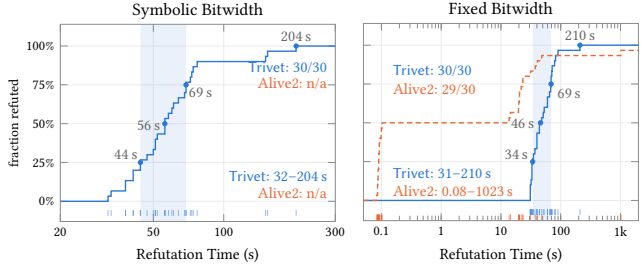
\begin{figure}[t]
  \centering
  \resizebox{\columnwidth}{!}{%
  \begin{tikzpicture}
    \begin{axis}[
      width=0.66\columnwidth, height=4.0cm, scale only axis,
      xmode=log, xmin=20, xmax=300, ymin=-0.13, ymax=1.14,
      title={\normalsize Symbolic Bitwidth},
      title style={yshift=-1.4ex, text height=1.8ex, text depth=0.4ex},
      xlabel={Refutation Time (s)}, ylabel={fraction refuted},
      xtick={20,50,100,300}, xticklabels={20,50,100,300},
      ytick={0,0.25,0.5,0.75,1},
      yticklabels={0\%,25\%,50\%,75\%,100\%},
      grid=major, grid style={draw=black!8},
      axis line style={draw=black!45}, tick style={draw=black!45},
      label style={font=\small}, tick label style={font=\footnotesize},
      ylabel style={font=\footnotesize,
        at={(axis description cs:-0.13,0.5)}, anchor=south},
      clip mode=individual,
    ]
      \addplot[draw=none, fill=L3blue, fill opacity=0.10]
        coordinates {(44,-0.13) (69,-0.13) (69,1.14) (44,1.14)} \closedcycle;
      \addplot[no marks, thick, color=L3blue] coordinates {(20,0.0000) (32,0.0000) (32,0.0333) (33,0.0333) (33,0.0667) (38,0.0667) (38,0.1000) (38,0.1000) (38,0.1333) (41,0.1333) (41,0.1667) (41,0.1667) (41,0.2000) (44,0.2000) (44,0.2333) (44,0.2333) (44,0.2667) (47,0.2667) (47,0.3000) (50,0.3000) (50,0.3333) (51,0.3333) (51,0.3667) (51,0.3667) (51,0.4000) (52,0.4000) (52,0.4333) (56,0.4333) (56,0.4667) (56,0.4667) (56,0.5000) (56,0.5000) (56,0.5333) (58,0.5333) (58,0.5667) (60,0.5667) (60,0.6000) (61,0.6000) (61,0.6333) (64,0.6333) (64,0.6667) (68,0.6667) (68,0.7000) (69,0.7000) (69,0.7333) (69,0.7333) (69,0.7667) (72,0.7667) (72,0.8000) (73,0.8000) (73,0.8333) (74,0.8333) (74,0.8667) (77,0.8667) (77,0.9000) (151,0.9000) (151,0.9333) (154,0.9333) (154,0.9667) (204,0.9667) (204,1.0000) (300,1.0000)};
      \addplot[only marks, mark=|, mark size=1.7pt, color=L3blue, opacity=0.5]
        coordinates {(32,-0.075) (33,-0.075) (38,-0.075) (38,-0.075) (41,-0.075) (41,-0.075) (44,-0.075) (44,-0.075) (47,-0.075) (50,-0.075) (51,-0.075) (51,-0.075) (52,-0.075) (56,-0.075) (56,-0.075) (56,-0.075) (58,-0.075) (60,-0.075) (61,-0.075) (64,-0.075) (68,-0.075) (69,-0.075) (69,-0.075) (72,-0.075) (73,-0.075) (74,-0.075) (77,-0.075) (151,-0.075) (154,-0.075) (204,-0.075)};
      \addplot[only marks, mark=*, mark size=1.25pt, color=L3blue]
        coordinates {(44,0.25) (56,0.50) (69,0.75) (204,1.00)};
      \node[font=\small,text=black!60,anchor=south east,inner sep=0.5pt,xshift=-1pt,yshift=1pt] at (axis cs:44,0.25) {44\,s};
      \node[font=\small,text=black!60,anchor=east,inner sep=0.5pt,xshift=-2pt,yshift=-1pt] at (axis cs:56,0.50) {\RqTwoSymMedian{}\,s};
      \node[font=\small,text=black!60,anchor=north west,inner sep=0.5pt,xshift=1pt,yshift=-1pt] at (axis cs:69,0.75) {69\,s};
      \node[font=\small,text=black!60,anchor=south,inner sep=0.5pt,yshift=2pt]
        at (axis cs:204,1.00) {204\,s};
      \node[font=\small,text=L3blue,anchor=east,inner sep=0.5pt]
        at (axis cs:280,0.86) {\proj: \RqTwoSymRefuted/\RqTwoSymCount};
      \node[font=\small,text=L3orange,anchor=east,inner sep=0.5pt]
        at (axis cs:280,0.76) {\alivetwo: n/a};
      \node[font=\small,text=L3orange,anchor=east,inner sep=0.5pt]
        at (axis cs:280,0.05) {\alivetwo: n/a};
      \node[font=\small,text=L3blue,anchor=east,inner sep=0.5pt]
        at (axis cs:280,0.16) {\proj: \RqTwoSymMin--\RqTwoSymMax{}\,s};
    \end{axis}
  \end{tikzpicture}\hspace{-0.2em}
  \begin{tikzpicture}
    \begin{axis}[
      width=0.66\columnwidth, height=4.0cm, scale only axis,
      xmode=log, xmin=0.05, xmax=2000, ymin=-0.13, ymax=1.14,
      title={\normalsize Fixed Bitwidth},
      title style={yshift=-1.4ex, text height=1.8ex, text depth=0.4ex},
      xlabel={Refutation Time (s)},
      xtick={0.1,1,10,100,1000}, xticklabels={0.1,1,10,100,1k},
      ytick={0,0.25,0.5,0.75,1}, yticklabels={,,,,},
      grid=major, grid style={draw=black!8},
      axis line style={draw=black!45}, tick style={draw=black!45},
      label style={font=\small}, tick label style={font=\footnotesize},
      clip mode=individual,
    ]
      \addplot[draw=none, fill=L3blue, fill opacity=0.10]
        coordinates {(34,-0.13) (69,-0.13) (69,1.14) (34,1.14)} \closedcycle;
      \addplot[no marks, thick, color=L3blue] coordinates {(0.05,0.0000) (31,0.0000) (31,0.0333) (31,0.0333) (31,0.0667) (31,0.0667) (31,0.1000) (32,0.1000) (32,0.1333) (33,0.1333) (33,0.1667) (33,0.1667) (33,0.2000) (33,0.2000) (33,0.2333) (34,0.2333) (34,0.2667) (35,0.2667) (35,0.3000) (37,0.3000) (37,0.3333) (40,0.3333) (40,0.3667) (40,0.3667) (40,0.4000) (40,0.4000) (40,0.4333) (44,0.4333) (44,0.4667) (44,0.4667) (44,0.5000) (48,0.5000) (48,0.5333) (51,0.5333) (51,0.5667) (53,0.5667) (53,0.6000) (60,0.6000) (60,0.6333) (60,0.6333) (60,0.6667) (65,0.6667) (65,0.7000) (68,0.7000) (68,0.7333) (69,0.7333) (69,0.7667) (69,0.7667) (69,0.8000) (71,0.8000) (71,0.8333) (72,0.8333) (72,0.8667) (79,0.8667) (79,0.9000) (82,0.9000) (82,0.9333) (90,0.9333) (90,0.9667) (210,0.9667) (210,1.0000) (2000,1.0000)};
      \addplot[no marks, thick, densely dashed, color=L3orange] coordinates {(0.05,0.0000) (0.083,0.0000) (0.083,0.0333) (0.086,0.0333) (0.086,0.0667) (0.086,0.0667) (0.086,0.1000) (0.087,0.1000) (0.087,0.1333) (0.088,0.1333) (0.088,0.1667) (0.088,0.1667) (0.088,0.2000) (0.090,0.2000) (0.090,0.2333) (0.090,0.2333) (0.090,0.2667) (0.090,0.2667) (0.090,0.3000) (0.092,0.3000) (0.092,0.3333) (0.092,0.3333) (0.092,0.3667) (0.093,0.3667) (0.093,0.4000) (0.099,0.4000) (0.099,0.4333) (0.100,0.4333) (0.100,0.4667) (0.103,0.4667) (0.103,0.5000) (13.698,0.5000) (13.698,0.5333) (14.465,0.5333) (14.465,0.5667) (19.629,0.5667) (19.629,0.6000) (19.890,0.6000) (19.890,0.6333) (20.344,0.6333) (20.344,0.6667) (20.478,0.6667) (20.478,0.7000) (21.818,0.7000) (21.818,0.7333) (23.259,0.7333) (23.259,0.7667) (23.310,0.7667) (23.310,0.8000) (31.263,0.8000) (31.263,0.8333) (37.494,0.8333) (37.494,0.8667) (40.711,0.8667) (40.711,0.9000) (48.165,0.9000) (48.165,0.9333) (1023.267,0.9333) (1023.267,0.9667) (2000,0.9667)};
      \addplot[only marks, mark=|, mark size=1.7pt, color=L3blue, opacity=0.5]
        coordinates {(31,-0.075) (31,-0.075) (31,-0.075) (32,-0.075) (33,-0.075) (33,-0.075) (33,-0.075) (34,-0.075) (35,-0.075) (37,-0.075) (40,-0.075) (40,-0.075) (40,-0.075) (44,-0.075) (44,-0.075) (48,-0.075) (51,-0.075) (53,-0.075) (60,-0.075) (60,-0.075) (65,-0.075) (68,-0.075) (69,-0.075) (69,-0.075) (71,-0.075) (72,-0.075) (79,-0.075) (82,-0.075) (90,-0.075) (210,-0.075)};
      \addplot[only marks, mark=|, mark size=1.7pt, color=L3orange, opacity=0.7]
        coordinates {(0.083,-0.110) (0.086,-0.110) (0.086,-0.110) (0.087,-0.110) (0.088,-0.110) (0.088,-0.110) (0.090,-0.110) (0.090,-0.110) (0.090,-0.110) (0.092,-0.110) (0.092,-0.110) (0.093,-0.110) (0.099,-0.110) (0.100,-0.110) (0.103,-0.110) (13.698,-0.110) (14.465,-0.110) (19.629,-0.110) (19.890,-0.110) (20.344,-0.110) (20.478,-0.110) (21.818,-0.110) (23.259,-0.110) (23.310,-0.110) (31.263,-0.110) (37.494,-0.110) (40.711,-0.110) (48.165,-0.110) (1023.267,-0.110)};
      \addplot[only marks, mark=*, mark size=1.25pt, color=L3blue]
        coordinates {(34,0.25) (46,0.50) (69,0.75) (210,1.00)};
      \node[font=\small,text=black!60,anchor=south east,inner sep=0.5pt,xshift=-2pt,yshift=1pt]
        at (axis cs:34,0.25) {34\,s};
      \node[font=\small,text=black!60,anchor=east,inner sep=0.5pt,xshift=-3pt]
        at (axis cs:46,0.50) {\RqTwoFixMedian{}\,s};
      \node[font=\small,text=black!60,anchor=north west,inner sep=0.5pt,xshift=3pt,yshift=-2pt]
        at (axis cs:69,0.75) {69\,s};
      \node[font=\small,text=black!60,anchor=south,inner sep=0.5pt,yshift=2pt]
        at (axis cs:210,1.00) {210\,s};
      \node[font=\small,text=L3blue,anchor=west,inner sep=0.5pt]
        at (axis cs:0.15,0.84) {\proj: \RqTwoFixRefuted/\RqTwoFixCount};
      \node[font=\small,text=L3orange,anchor=west,inner sep=0.5pt]
        at (axis cs:0.15,0.72) {\alivetwo: \AliveCexRefuted/\AliveCexCount};
      \node[font=\small,text=L3blue,anchor=west,inner sep=0.5pt]
        at (axis cs:0.15,0.18) {\proj: \RqTwoFixMin--\RqTwoFixMax{}\,s};
      \node[font=\small,text=L3orange,anchor=west,inner sep=0.5pt]
        at (axis cs:0.15,0.06) {\alivetwo: \AliveCexMin--\AliveCexMax{}\,s};
    \end{axis}
  \end{tikzpicture}
  }
  \caption{Refutation time distributions for \rqtwo (log scale).}
  \label{fig:rq2-dist}
\end{figure}

In \rqtwo, we investigate whether \proj can refute real-world invalid
\llvm transformations and compare its refutation capability with \alivetwo.
\Cref{tab:rq2} summarizes the aggregate results, while
\Cref{fig:rq2-dist} shows the corresponding refutation-time distributions.

\subsubsection{Experimental Results}

\proj produces certified refutations
for all \RqTwoCount invalid transformations, with a mean refutation time of
\RqTwoMean{}\,s.  Candidate generation accounts for \RqTwoProposeMean{}\,s
(\RqTwoProposeShare), while scaffold-based certification accounts for
\RqTwoCertifyMean{}\,s (\RqTwoCertifyShare).
For the symbolic-bitwidth transformations, \proj refutes all \RqTwoSymCount cases in
\RqTwoSymMin--\RqTwoSymMax{}\,s, with a median
of \RqTwoSymMedian{}\,s and a mean of \RqTwoSymMean{}\,s.  For each case,
the \llm proposes a counterexample candidate containing concrete bitwidths
and input values, which the scaffold automatically certifies.
For the fixed-bitwidth transformations, \proj refutes all
\RqTwoFixCount cases in \RqTwoFixMin--\RqTwoFixMax{}\,s, with a median of
\RqTwoFixMedian{}\,s and a mean of \RqTwoFixMean{}\,s.
In contrast, \alivetwo refutes \AliveCexRefuted and
times out on the remaining one.
Successful runs take \AliveCexMin--\AliveCexMax{}\,s, with a median of
\AliveCexMedian{}\,s and a mean of \AliveCexMean{}\,s.

\subsubsection{Result Analysis}

The symbolic-bitwidth results show that \proj can refute a width-parametric
transformation by certifying a counterexample that instantiates both the bitwidth and
inputs, extending coverage beyond \alivetwo's scope.  For fixed bitwidths,
\alivetwo is faster on most cases but still exhibits a pronounced long tail and
leaves one case unresolved after two hours, consistent with its behavior on
valid transformations in \Cref{sec:rq1}.  In contrast, \proj has more
concentrated runtimes and refutes every case.

\subsection{\rqthree: Loop-Containing Transformations}
\label{sec:rq3}

\Cref{tab:rq3} summarizes the experimental results for the \RqThreeEvaluatedCount instances.
\proj provides refinement proofs for all valid cases at both
fixed and symbolic bitwidths. The symbolic proofs establish
correctness for all widths and inputs, with median and mean
proof-generation times of \RqThreeValidSymMedian~s and \RqThreeValidSymMean~s.
The fixed-width runs have corresponding times of
\RqThreeValidFixMedian~s and \RqThreeValidFixMean~s.
Fixed-width refutation certifies
\RqThreeInvalidFixRefuted/\RqThreeInvalidCaseCountFixed invalid cases,
with a median time of \RqThreeInvalidFixMedian~s.
Symbolic-width refutation certifies all
\RqThreeInvalidEvalCountSymbolic invalid cases,
with a median of
\RqThreeInvalidSymMedian~s. Each symbolic counterexample specifies
concrete inputs and a bitwidth.
We omit \alivetwo because it does not support symbolic bitwidths and
establishes correctness only within a user-configurable unrolling bound,
rather than proving the transformation correct for all loop iteration counts~\cite{alivetwo}.

For the unresolved invalid transformation, \proj exhausted
its counterexample candidate attempts after 306~s without finding
a certified counterexample.
With \mbox{$x,n,a_i:\texttt{i32}$}, the transformation replaces
a loop with a multiplication:
\[
a_0=0,\quad a_{i+1}=a_i+x,\quad
\operatorname{return}a_n \to \operatorname{return}(x\times n).
\]
For \texttt{x=poison}, \texttt{n=0}, the source executes no loop iterations,
leaving the accumulator at its initial value of zero.
The target instead evaluates \texttt{x * n}, which propagates poison.
Although the arithmetic identity holds for defined operands, the
rewrite introduces a dependence on \texttt{x} into the source's
otherwise constant return value when \texttt{n=0}.

\begin{table}[t]
  \centering
  \caption{\rqthree results for loop-containing transformations.
  }
  \label{tab:rq3}
  \footnotesize
  \renewcommand{\arraystretch}{1.0}
  \setlength{\tabcolsep}{5pt}
  \begin{tabular}{@{}llrrrr@{}}
    \toprule
    \textbf{Bitwidth} & \textbf{Task} & \textbf{$N$} & \textbf{Succeeded}
                      & \textbf{Median\,(s)} & \textbf{Mean\,(s)} \\
    \midrule

    Symbolic & Validation & \RqThreeValidCaseCountSymbolic
             & \RqThreeValidSymProved & \RqThreeValidSymMedian
             & \RqThreeValidSymMean \\
    Fixed    & Validation & \RqThreeValidCaseCountFixed
             & \RqThreeValidFixProved & \RqThreeValidFixMedian
             & \RqThreeValidFixMean \\
    Symbolic & Refutation & \RqThreeInvalidEvalCountSymbolic
             & \RqThreeInvalidSymRefuted & \RqThreeInvalidSymMedian
             & \RqThreeInvalidSymMean \\
    Fixed    & Refutation & \RqThreeInvalidCaseCountFixed
             & \RqThreeInvalidFixRefuted & \RqThreeInvalidFixMedian
             & \RqThreeInvalidFixMean \\
    \bottomrule
  \end{tabular}
\end{table}

\subsection{\rqfour: Ablation Study}
\label{sec:rq4}

In \rqfour, we assess the contribution of the scaffold by comparing
\proj with its unscaffolded variant \projminus, which requires
the \llm to generate complete \lean proofs from scratch.
Keeping all other experimental settings fixed and evaluating
on the RQ1--RQ3 datasets, we measure proof success rate, runtime, and
monetary cost in the ablation study.

\begin{table}[t]
  \centering
  \caption{Scaffold ablation results.
  Mean runtimes are in seconds for successful cases in
  both configurations. The column $\Delta$ is the cost
  reduction from \projminus to \proj.
  }
  \label{tab:rq4-ablation}
  \footnotesize
  \renewcommand{\arraystretch}{1}
  \setlength{\aboverulesep}{0.3ex}
  \setlength{\belowrulesep}{0.4ex}
  \setlength{\tabcolsep}{3pt}
  \resizebox{\columnwidth}{!}{%
  \begin{tabular}{@{}lrrrrrrrrr@{}}
    \toprule
    & & \multicolumn{2}{c}{\textbf{Successful}}
      & \multicolumn{3}{c}{\textbf{Mean Runtime (s)}}
      & \multicolumn{3}{c}{\textbf{Mean Cost (USD)}} \\
    \cmidrule(lr){3-4}\cmidrule(lr){5-7}\cmidrule(lr){8-10}
    \textbf{Bitwidth} & \textbf{$N$} & \textbf{\projminus} & \textbf{\proj}
                     & \textbf{\projminus} & \textbf{\proj}
                     & \textbf{Speedup} & \textbf{\projminus} & \textbf{\proj}
                     & \textbf{$\Delta$} \\
    \midrule
    \multicolumn{10}{c}{\emph{Refinement proofs (loop-free)}} \\
    Symbolic          & \AblSymCount & \AblBaseSymSolved & \AblScafSymSolved
                      & \AblPairSymBaseMean & \AblPairSymScafMean
                      & \AblPairSymSpeedup
                      & \$\AblCostSymBaseMean & \$\AblCostSymScafMean
                      & \AblCostSymReduction \\
    Fixed             & \AblFixCount & \AblBaseFixSolved & \AblScafFixSolved
                      & \AblPairFixBaseMean & \AblPairFixScafMean
                      & \AblPairFixSpeedup
                      & \$\AblCostFixBaseMean & \$\AblCostFixScafMean
                      & \AblCostFixReduction \\
    Total             & \AblBaseCount & \AblBaseSolved & \AblScafSolved
                      & \AblPairBaseMean & \AblPairScafMean & \AblPairSpeedup
                      & \$\AblCostBaseMean & \$\AblCostScafMean
                      & \AblCostReduction \\
    \midrule
    \multicolumn{10}{c}{\emph{Counterexample proofs (loop-free)}} \\
    Symbolic          & \AblCexSymCount & \AblCexBaseSymRefuted
                      & \AblCexScafSymRefuted
                      & \AblCexPairSymBaseMean & \AblCexPairSymScafMean
                      & \AblCexPairSymSpeedup
                      & \$\AblCexCostSymBaseMean & \$\AblCexCostSymScafMean
                      & \AblCexCostSymReduction \\
    Fixed             & \AblCexFixCount & \AblCexBaseFixRefuted
                      & \AblCexScafFixRefuted
                      & \AblCexPairFixBaseMean & \AblCexPairFixScafMean
                      & \AblCexPairFixSpeedup
                      & \$\AblCexCostFixBaseMean & \$\AblCexCostFixScafMean
                      & \AblCexCostFixReduction \\
    Total             & \AblCexCount & \AblCexBaseRefuted
                      & \AblCexScafRefuted
                      & \AblCexPairBaseMean & \AblCexPairScafMean
                      & \AblCexPairSpeedup
                      & \$\AblCexCostBaseMean & \$\AblCexCostScafMean
                      & \AblCexCostReduction \\
    \midrule
    \multicolumn{10}{c}{\emph{Refinement proofs (loop-containing)}} \\
    Symbolic          & \AblLoopSymCount & \AblLoopBaseSymSolved
                      & \AblLoopScafSymSolved
                      & \AblLoopPairSymBaseMean & \AblLoopPairSymScafMean
                      & \AblLoopPairSymSpeedup
                      & \$\AblLoopCostSymBaseMean & \$\AblLoopCostSymScafMean
                      & \AblLoopCostSymReduction \\
    Fixed             & \AblLoopFixCount & \AblLoopBaseFixSolved
                      & \AblLoopScafFixSolved
                      & \AblLoopPairFixBaseMean & \AblLoopPairFixScafMean
                      & \AblLoopPairFixSpeedup
                      & \$\AblLoopCostFixBaseMean & \$\AblLoopCostFixScafMean
                      & \AblLoopCostFixReduction \\
    Total             & \AblLoopCount & \AblLoopBaseSolved
                      & \AblLoopScafSolved
                      & \AblLoopPairBaseMean & \AblLoopPairScafMean
                      & \AblLoopPairSpeedup
                      & \$\AblLoopCostBaseMean & \$\AblLoopCostScafMean
                      & \AblLoopCostReduction \\
    \midrule
    \multicolumn{10}{c}{\emph{Counterexample proofs (loop-containing)}} \\
    Symbolic          & \AblLoopCexSymCount & \AblLoopCexBaseSymRefuted
                      & \AblLoopCexScafSymRefuted
                      & \AblLoopCexPairSymBaseMean & \AblLoopCexPairSymScafMean
                      & \AblLoopCexPairSymSpeedup
                      & \$\AblLoopCexCostSymBaseMean & \$\AblLoopCexCostSymScafMean
                      & \AblLoopCexCostSymReduction \\
    Fixed             & \AblLoopCexFixCount & \AblLoopCexBaseFixRefuted
                      & \AblLoopCexScafFixRefuted
                      & \AblLoopCexPairFixBaseMean & \AblLoopCexPairFixScafMean
                      & \AblLoopCexPairFixSpeedup
                      & \$\AblLoopCexCostFixBaseMean & \$\AblLoopCexCostFixScafMean
                      & \AblLoopCexCostFixReduction \\
    Total             & \AblLoopCexCount & \AblLoopCexBaseRefuted
                      & \AblLoopCexScafRefuted
                      & \AblLoopCexPairBaseMean & \AblLoopCexPairScafMean
                      & \AblLoopCexPairSpeedup
                      & \$\AblLoopCexCostBaseMean & \$\AblLoopCexCostScafMean
                      & \AblLoopCexCostReduction \\
    \bottomrule
  \end{tabular}
  }
\end{table}

\subsubsection{Success Rate and Runtime}
\label{sec:rq4-performance}

For loop-free transformations, \Cref{tab:rq4-ablation} shows that
\proj completes all refinement and counterexample proofs.
In contrast, \projminus succeeds on
\AblBaseSolved{} of \AblBaseCount{} refinement cases and
\AblCexBaseRefuted{} of \AblCexCount{} counterexample cases.
Among cases successfully completed by both configurations, \proj is
faster in every case. The speedups in mean runtime are
\AblPairSpeedup{} for refinement and \AblCexPairSpeedup{} for
counterexamples.
For loop-containing transformations, \projminus and \proj prove
\AblLoopBaseSolved{} and \AblLoopScafSolved{} of \AblLoopCount{}
valid instances, respectively, and refute \AblLoopCexBaseRefuted{}
and \AblLoopCexScafRefuted{} of \AblLoopCexCount{} invalid instances.
Among cases solved by both configurations, the ratios of baseline to
scaffolded mean runtime are \AblLoopPairSpeedup{} for refinement and
\AblLoopCexPairSpeedup{} for counterexamples.

These results are consistent with scaffolding reducing proof
generation overhead.
We observe that generating proofs from scratch can be inefficient
due to extensive tactic exploration and repeated proof attempts,
with loose proof structure and diverse \ub cases adding further complexity.
In contrast, the scaffold guides the \llm through well-defined subgoals, yielding
proofs with a clearer logical structure.

\subsubsection{Monetary Cost}
\label{sec:rq4-cost}

We report per-transformation proof-generation costs based on
OpenAI's pricing as of September 2026~\cite{openai_api_pricing}.
For loop-free transformations, mean costs decrease from
\$\AblCostBaseMean{} to \$\AblCostScafMean{} for the \AblBaseSolved{}
valid cases and from \$\AblCexCostBaseMean{} to \$\AblCexCostScafMean{}
for the \AblCexBaseRefuted{} invalid cases, corresponding to reductions
of \AblCostReduction{} and \AblCexCostReduction{}, respectively.
For loop-containing transformations with complete usage records,
mean costs decrease from \$\AblLoopCostBaseMean{} to
\$\AblLoopCostScafMean{} for refinement proofs and from
\$\AblLoopCexCostBaseMean{} to \$\AblLoopCexCostScafMean{}
for counterexample proofs.

%% file: related-work.tex
We review the following three areas related to \proj.

\myparagraph{Translation Validation}
Translation validation checks individual compiler runs or transformations
rather than verifying the compiler implementation~\cite{tv,opttv,tvoc}.
Equality-based approaches use equality saturation or normalized value graphs
to check \llvm and \mlir transformations~\cite{equalitybbasedtv,vgtv,HEC}.
Alive~\cite{aliveone} and \alivetwo~\cite{alivetwo} provide \smt-based
refinement checking for \llvm.  AliveInLean follows this design by
verifying its condition generator in \lean but still relies
on an \smt solver, whereas Crellvm checks compiler-emitted
proofs with a Coq-verified validator~\cite{aliveinlean,crellvm}.
\proj instead uses an \llm to synthesize
per-transformation \lean proofs checked by the kernel.

\myparagraph{\llms for Compilers}
Recent work applies \llms to compiler optimization and debugging.
For optimization development, LPO~\cite{lpo}
discovers formally verified peephole optimizations, while LPG~\cite{lpg} generalizes
concrete optimizations into reusable rules.  Recent studies
evaluate agents implementing \llvm
optimizations~\cite{xu2026codingagentsimplementmissed,guan2026understandingagentbasedpatchingcompiler}.
For compiler debugging, LPR and subsequent agentic work minimize
bug-triggering programs~\cite{LPR,agenticreduction}.
By contrast, \proj uses an \llm to construct
correctness or counterexample proofs for a given transformation, with \lean checking the result.

\myparagraph{\llms for Interactive Theorem Proving}
\llms have been used to automate proof construction in interactive theorem
provers.  LeanDojo and Rango use retrieval augmentation for \lean and \coq,
respectively~\cite{leandojo,Rango}.  COPRA combines stateful search with prover
feedback, while PALM iteratively repairs generated
proofs~\cite{COPRA,proofautomation}.
\proj applies \llm-based proof synthesis to \llvm translation
validation, using a domain-specific scaffold to produce
machine-checked proofs in \lean.

%% file: conclusion.tex
This paper presents \proj, a framework integrating \llms and \lean
for automated \llvm translation validation.
By combining deterministic scaffolds, \llm-guided proof synthesis,
and \lean kernel checking, \proj produces refinement and counterexample
proofs for fixed and symbolic bitwidths, covering loop-free transformations
and a restricted class of loop-containing ones.
Our evaluation shows that \proj validates transformations beyond the practical
reach of \alivetwo, the state-of-the-art \smt-based validator, while
scaffolding improves proof completion and efficiency.